\documentclass[sn-nature]{sn-jnl}

\usepackage{graphicx}%
\usepackage{multirow}%
\usepackage{amsmath,amssymb,amsfonts}%
\usepackage{amsthm}%
\usepackage{mathrsfs}%
\usepackage[title]{appendix}%
\usepackage{xcolor}%
\usepackage{textcomp}%
\usepackage{manyfoot}%
\usepackage{booktabs}%
\usepackage{algorithm}%
\usepackage{algorithmicx}%
\usepackage{algpseudocode}%
\usepackage{listings}%
\usepackage{comment}
\usepackage{ulem}

\usepackage[colorinlistoftodos]{todonotes}

\begin{document}

\title[Article Title]{Human spatial dynamics for electricity demand forecasting: the case of France during the 2022 energy crisis}


\author*[1,2]{\fnm{Nathan} \sur{Doumèche}}\email{nathan.doumeche@sorbonne-universite.fr}

\author[2]{\fnm{Yann} \sur{Allioux}}\email{yann.allioux@edf.fr}

\author[2]{\fnm{Yannig} \sur{Goude}}\email{yannig.goude@edf.fr}

\author[3]{\fnm{Stefania} \sur{Rubrichi}}\email{stefania.rubrichi@orange.com}

\affil*[1]{\orgdiv{LPSM}, \orgname{Sorbonne University}, \orgaddress{\street{4 place Jussieu}, \city{Paris}, \postcode{75005}, \country{France}}}

\affil[2]{\orgdiv{OSIRIS}, \orgname{EDF R$\&$D }, \orgaddress{\street{ Bd Gaspard Monge}, \city{Saclay}, \postcode{91120}, \country{France}}}

\affil[3]{\orgdiv{SENSE lab}, \orgname{Orange Innovation}, \orgaddress{\street{46 Av. de la République}, \city{Châtillon}, \postcode{92320}, \country{France}}}


\abstract{
Accurate electricity demand forecasting is crucial to meet energy security and efficiency, especially when relying on intermittent renewable energy sources. Recently, massive savings have been observed in Europe, following an unprecedented global energy crisis. 
However, assessing the impact of such crisis and of government incentives on electricity consumption behaviour is challenging. 
Moreover, standard statistical models based on meteorological and calendar data have difficulty adapting to such brutal changes. 
Here, we show that mobility indices based on  
mobile network data significantly improve the performance of the state-of-the-art models in electricity demand forecasting during the sobriety period. We start by documenting the drop in the French electricity consumption during the winter of 2022-2023. 
We then show how our mobile network data captures work dynamics and how adding these mobility indices outperforms the state-of-the-art during this atypical period. 
Our results 
characterise the effect of work behaviours on the electricity demand.
  }
 


\maketitle


Energy is at the very core of modern economies and politics, powering industry, transport, residential use, and agriculture 
\citep{shove2014energy}. 
Over the past two years, Europe has experienced a major energy crisis, with energy prices reaching levels not seen in decades \citep{ferrani2023the}. Prices began to rise rapidly in the summer of 2021 as the global economy picked up following the easing of COVID-19 restrictions. Subsequently, the war in Ukraine led to a significant reduction in gas supplies, pushing gas prices even higher \citep{ruhnau2023natural}. 
In this context, the European Union adopted the Council Regulation 2022/1854 in October 2022 \cite{european2022council}. 
This regulation established a series of emergency measures to mitigate the effects of such a crisis, mainly by reducing the electricity demand with a binding reduction target of 5 \% during peak hours.
In France, in particular, where a significant proportion of the nuclear plants were also offline \citep{rteFeb2023}, 
the government called for a voluntary mobilisation to reduce energy consumption by 10\% over two years and launched its own energy sobriety plan \cite{gouvernement2022plan}.
Various media have documented a drop in the French electricity demand in the winter of 2022-2023 \citep{technique2022consommation, thenyt2022asrussia, lemonde2023labaisse}.
Energy saving is also part of France's long-term policy of ecological transition and energy sovereignty.
Indeed, the impact of the energy sector on climate change is forcing the adaptation of consumption patterns, which is fueling a growing interest in energy savings and the transition to sustainable energy sources \citep{abdelaziz2011a, rockstrom2017a, hoegh2019the, demaere2020the}.
In France, electricity is one of the most important components of the energy mix, accounting for 25\% of French final energy consumption, and the French Ecological Transition Plan is based on a massive electrification driven by decarbonised energy coupled with energy savings \citep{omar2022decarbonizing, rte2022}. 
While adapting human behaviour (e.g. by encouraging remote working) has been identified as an important axis of the sobriety plan, a better understanding of how this relates to energy savings is crucial for energy planning.

Recently, artificial intelligence has been recognised as a powerful tool to support the mitigation of greenhouse gas emissions and tackle climate change \citep{rolnick2022tackling}. 
In particular, machine learning techniques have been applied to electricity load forecasting to ensure the balance of the electricity grid \citep{pinheiro2023short} and to reduce electricity waste. 
As electricity storage capacity is limited and expensive, electricity supply must match demand at all times. 
As a result, electricity load forecasting at different forecast horizons has attracted increasing interest over the last few years \citep{hong2020energy}. 
This article focuses on the so-called short-term load forecasting, or 24-hour ahead load forecasting, which is particularly relevant for operational usages in industry and the electricity market  \citep{nti2020review, hammad2020methods}. 
We address this problem both in terms of feature selection and model design.
Most state-of-the-art models rely on historical data of past electricity loads, calendar data such as holidays or the position of the day in the week, and meteorological data such as temperature and humidity \citep{nti2020review}. 
However, such data cannot accurately account for the complex human behaviours that affect the variability of energy demand, such as holidays or remote working. 
As a result, traditional models have struggled to account for brutal societal events such as the COVID-19 lockdowns, or energy savings following economic, geopolitical, and environmental crises \citep{obst2021adaptative}.
New data capturing consumption behaviours is needed to better model the electricity demand. 
Over the last decades, datasets generated from cell phones networks, location-based services (LBS), and remote sensors in general  have emerged to sense human behaviours \citep{blondel_understanding_2015}.
Indeed, geolocation from mobile phones makes it possible to precisely characterise human flows \citep{deville_dynamic_2014, blumenstock_predicting_2015, lorenzo_allaboard_2016}. 
For example, such data have been used to study disease propagation \citep{bengtsson_improved_2011, blumenstock_inferring_2012, rubrichi_comparison_2018, pullano_evaluating_2020}, traffic
\citep{xu2021understanding}, the impact of human activities on biodiversity \citep{filazzola2022using}, and water consumption  \citep{terroso2021human, smolak2020applying}. 
In terms of day-ahead load forecasting, mobility data from SafeGraph, Google, and Apple mobility reports were strongly correlated with electricity load drops in the US during the COVID-19 outbreaks \citep{chen2020using, ruan2020cross}, as well as in Ireland \citep{zarbakhsh2022human} and in France \citep{antoniadis2021hierarchical}. 
Although these datasets are very informative about the activity of urban areas, e.g. in retail stores or train stations, they were not intended to precisely account for human presence or flows. Indeed, there is an intrinsic bias in the data collection, corresponding for example to the bias of using a specific application, which causes the need to adjust the indices.

Thus, the originality of this paper relies on the use of the adjusted high-quality human presence data provided by the mobile network operator Orange, which represents between $30 \%$ and $40 \%$ of the French telephony market share, to model electricity demand during the sobriety period in France in 2022-2023 \cite{fluxVision}.
This dataset is based on  traffic volume measurements collected continuously and unobtrusively at the mobile network level, whereas most LBS data depends on users explicitly agreeing to share their location with specific applications. 
As a result, our mobile network based signal is very stable and suitable for census \cite{Levy2023who}.
In this article, we start by characterising electricity savings during the sobriety period in France. 
We then show that our mobility data from mobile networks are correlated with other well-known socio-economic indices that capture the spatial dynamics of populations. Furthermore, we show that models using mobility data outperforms by an order of magnitude of 10\% the state-of-the-art in electricity demand forecasting.
Finally, we show that 
our work index (see Section \ref{sec:mob_dataset}) has a strong and distinctive effect on the electricity demand, making sense of the drops in electricity demand during holidays. 
Other human spatial dynamics indices, such as residency and tourism at national levels, did not prove to have a significant effect on the national electricity demand.


\section{Results}
\subsection{Quantifying electricity savings} 
\label{sec:energy_savings}
To quantify electricity savings, the effect of temperature and time seasonality must be removed from the French electricity demand.
This effect, which we denote by $\mathrm{L}\widehat{\mathrm{oa}} \mathrm{d}$, is estimated by a Generalized Additive Model (GAM).
\begin{figure}
    \centering
    \caption{\textbf{Electricity demand corrected for the effects of temperature and annual seasonality.}}
    \includegraphics[width = 0.49\textwidth]{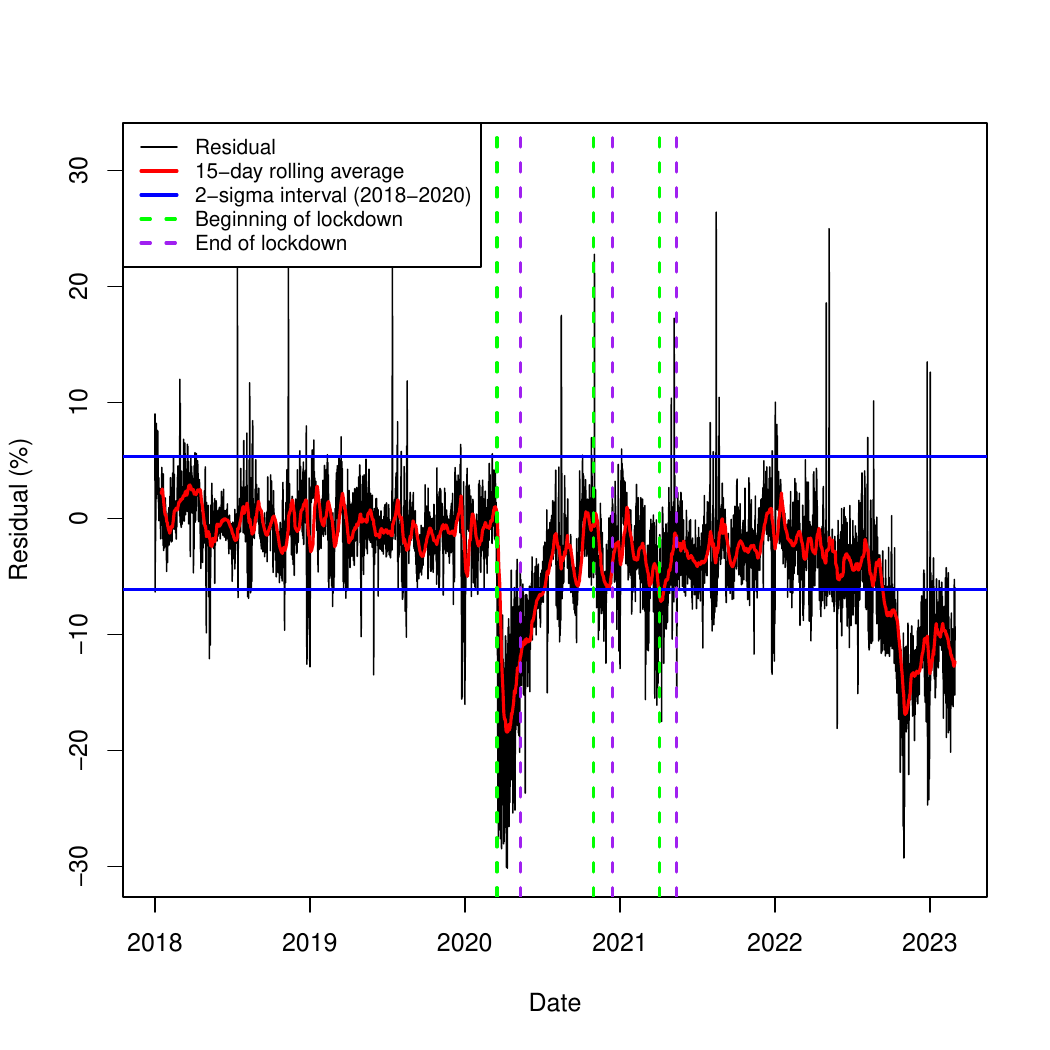}
    \includegraphics[width = 0.49\textwidth]{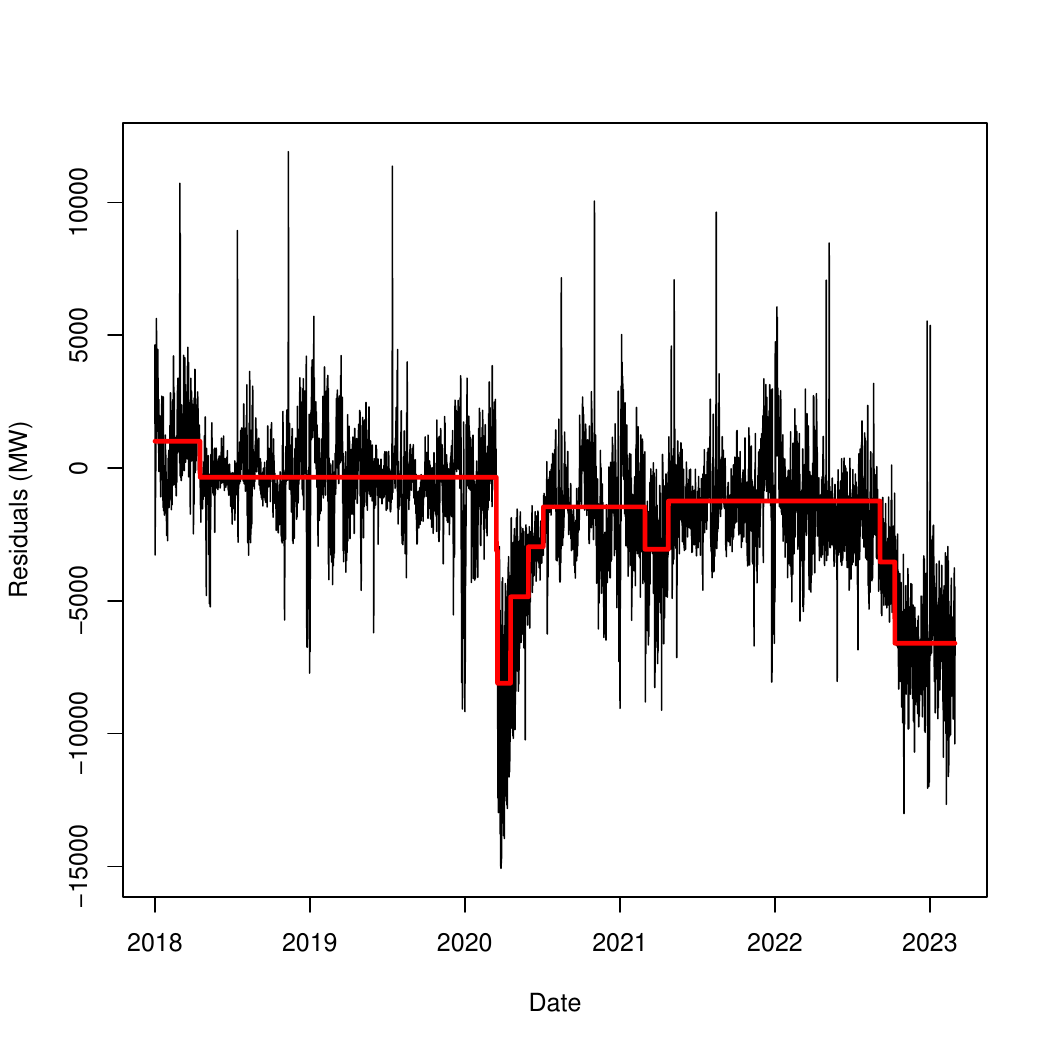}
     \\  
     \flushleft
     {\small
     \textbf{Left:} Descriptive statistics of the residuals.  \textbf{Right:} The 10 most important change points are represented by a change in the red line. The red line is the mean of the residuals between the change points.}
    \label{fig:obst}
\end{figure}
Figure \ref{fig:obst} shows the residuals $\mathrm{res} = \mathrm{Load} - \mathrm{L}\widehat{\mathrm{oa}} \mathrm{d}$, where Load is the actual value of the electricity demand.
This GAM is trained from 2014-01-01 to 2018-01-01. The residuals are then evaluated from 2018-01-01 to 2023-03-01.
Therefore, residuals measure the gap between the electricity demand at a given time and the expected demand with respect to its time and temperature dependency between 2014 and 2018. Negative residuals correspond to electricity savings.  In Figure \ref{fig:obst} (left), the blue lines represent the 2-$\sigma$ variations over the period spanning from 2018-01-01 to 2020-01-01. They correspond to the typical variations of the electricity demand around its expected value given the temperature and the position in the year.
The holidays deviate strongly from the expected trend and correspond to the peaks in the residuals.
Note that the 15-day rolling average in red only leaves this confidence interval during the lockdowns and the winter 2022-2023 sobriety period. 
This means that, during these events, the French electricity load is significantly lower than its normal values. To quantify these changes in the electricity demand, we run a change point analysis using the \textit{changepoint} package \citep{killik2014changepoint, Aminikhanghahi2017a}. 
More specifically, we run the binary segmentation algorithm, which detects and orders the changes in the mean of the residuals.
As a result, the two most important change points of the 2018-2023 period are the beginning of the sobriety period (2022-10-10) and the first COVID-19 lockdown (2020-03-15). During the sobriety period running from 2022-10-10 to 2023-03-01, the residuals have a mean of -10.6 $\%$. This result is close to the assessment made by the French Transmission System Operator's estimate of a 9$\%$ decrease in consumption during the winter of 2022-2023 \cite{rte2023}.
Figure \ref{fig:obst} (right) shows the 10 most important change points of the 2018-2023 period in the residuals and presents the mean gap during the corresponding periods. Notice the transition running from 2022-09-05 to 2022-10-10 between the stable regime of 2021-2022 and the sobriety period. 
These results prove that there was a significant drop of 10.6 $\%$ in the French electricity demand during the sobriety period, running from 2022-10-10 to 2023-03-01, that temperature and calendar data are not sufficient to  explain accurately.

\subsection{Mobile phone data and work dynamics}
\begin{figure}
    \centering
    \caption{\textbf{Comparison of work indices.}}
    \includegraphics[width = 0.6\textwidth]{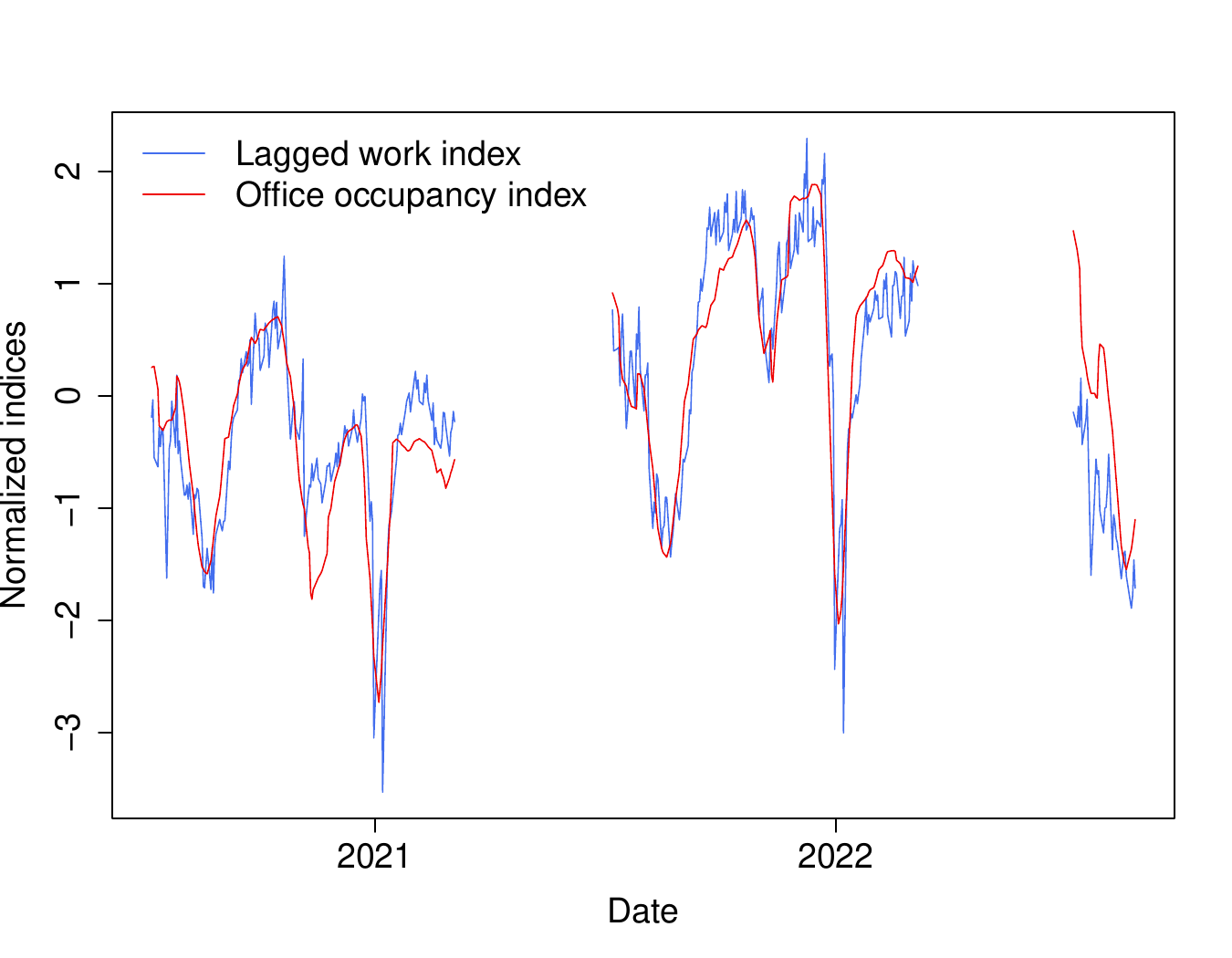}
     \flushleft
     {\small This figure shows the mobile network-based work index and the Normalcy office occupancy one. The work index in blue is lagged by 7 days. Weekends and bank holidays are excluded. Both indices have been normalised, i.e., subtracted their empirical mean and divided by their empirical standard deviation. The mobile network dataset only covers the period from July to March each year.} 
    \label{fig:normalcy}
\end{figure}

Here, we show how our dataset of aggregated mobile phone data efficiently captures the spatial dimension of social dynamics. 
In particular, we focus on its ability to track changes in the distribution of work-related human presence (hereafter \textit{work index}) over time.  
To investigate the ability of our dataset to characterise such behaviour, we used the \textit{office occupancy} index from The Economist's Normalcy index \citep{theeconomist2022theglobal}. 
This index was developed during the COVID-19 pandemic to evaluate the impact of the pandemic and government policies on human behaviour.
It tracks eight variables (sports attendance, time at home, traffic congestion, retail footfall, office occupancy, flights, film box office, and public transport) at national level,  which are openly available at 
\url{https://github.com/TheEconomist/normalcy-index-data}. 
The \textit{office occupancy} index is derived from the Google COVID-19 Community Mobility Reports, which are no longer being updated as of mid-October 2022.
As illustrated by Figure \ref{fig:normalcy}, the office occupancy variable has an 87 \% correlation with the 7-day lagged mobile phone index when excluding weekends and bank holidays. 
Moreover, as detailed in Appendix \ref{sec:data}, our mobile phone index carries more information, because it captures the reduction in office occupancy during weekends and holidays, and because it is seven days ahead of the normalcy index. 

Indeed, holidays are known to  have a significant impact on  electricity demand, while their effect is difficult to evaluate. 
This often leads to analyse regular days and of holidays separately \cite{Krstonijevic2022adaptive}.
In the appendices, we demonstrate how the tourism trends are related to another mobile phone index from the same dataset.

\subsection{Load forecasting with mobility data }
\label{sec:benchmark}
The purpose of this paragraph is to measure the benefits of incorporating mobility data into state-of-the-art load forecasting techniques (see Appendix \ref{app:Benchmark} for a more complete description of the models). 
In this field, the state-of-the-art is generally divided into three classes of forecasts \cite{Singh2012Load, wang2023benchmarks}: statistical models that approximate electricity demand by simple relationships between explanatory variables, data assimilation techniques that update a model by using recent observations, and machine learning methods that are less explainable but model-free.  
Here, we will focus on the state-of-the-art in the French load forecasting.
To evaluate the benefits of using mobility data to forecast the French national electricity load, we run a benchmark on the sobriety period, i.e., evaluated from 2022-01-09 to 2023-02-28. The training period spans from 2013-01-08 to 2022-01-09. Results are presented in Table \ref{table_score_target_agg2} in terms of Root Mean Square Error (RMSE) and Mean Absolute Percentage Error (MAPE). 
Bold values highlights the best forecasts in each category. 
Overall, they show that incorporating mobility data improves the  performance of the best forecast (aggregation of experts) by about 15\% in RMSE and 10\% in MAPE. 
For a full description of the models and metrics, please refer to the Methods section.
These gains are significant, because they leave the confidence intervals obtained by bootstrapping.
Overall, adding mobility data improves the performance of all models by an order of magnitude of $10 \%$ and the ranking of the models is consistent with recent studies \cite{obst2021adaptative, vilmarest2022state}.
Notice how the time series bootstrap improves the performance of the random forests without mobility data, confirming the results of \cite{gohery2023random}, but how this is not the case when adding mobility data.
\begin{table}[ht]
\centering
\caption{\textbf{Benchmark with and without mobility data.}} 
\begin{tabular*}{\textwidth}{@{\extracolsep\fill}lcccc}
  \toprule
  & \multicolumn{2}{@{}c@{}}{ Without mobility data  } & \multicolumn{2}{@{}c@{}}{With mobility data } \\\cmidrule{2-3}\cmidrule{4-5}%
 & RMSE (GW) & MAPE (\%)  &  RMSE (GW) & MAPE (\%) \\
  \midrule
  \textit{Statistical model} &&&&\\
  Persistence (1 day) & 4.0 $\pm$ 0.2 & 5.5 $\pm$ 0.3 & N.A. & N.A.\\
  SARIMA  &  2.4  $\pm$  0.2   & \textbf{3.1}  $\pm$ 0.2 & N.A. & N.A. \\
  GAM & 2.3 $\pm$ 0.1 & 3.5 $\pm$ 0.2   & \textbf{2.17} $\pm$ 0.08  & 3.3 $\pm$ 0.1  \\
  \midrule
    \textit{Data assimilation technique}\\
  Static Kalman filter & 2.1 $\pm$ 0.1 & 3.1 $\pm$ 0.2   &  1.72 $\pm$ 0.08 & 2.5  $\pm$ 0.1 \\
  Dynamic Kalman filter & 1.4 $\pm$ 0.1 & 1.9 $\pm$ 0.1   & 1.20 $\pm$ 0.08 & 1.7  $\pm$ 0.1  \\
    Viking & 1.5 $\pm$ 0.1 & 1.8 $\pm$ 0.1 &  1.24 $\pm$ 0.07 & 1.7  $\pm$ 0.1  \\
    Aggregation of experts & 1.4 $\pm$ 0.1 & 1.8 $\pm$ 0.1 & \textbf{1.16} $\pm$ 0.07 & \textbf{1.6}  $\pm$ 0.1  \\
    \midrule
    \textit{Machine learning}\\
    GAM boosting & 2.6 $\pm$ 0.2 & 3.7 $\pm$ 0.2 & 2.4 $\pm$ 0.1 & 3.5 $\pm$ 0.2 \\
    Random forests &  2.5 $\pm$ 0.2& 3.5 $\pm$ 0.2& \textbf{2.0} $\pm$ 0.1& \textbf{2.7} $\pm$ 0.2\\
    Random forests + bootstrap & 2.2 $\pm$ 0.2 & 3.0 $\pm$ 0.2 & \textbf{2.0} $ \pm$ 0.1 & \textbf{2.7} $\pm$ 0.2\\
   \bottomrule
\end{tabular*}
\label{table_score_target_agg2}
\end{table}
Finally, holidays are known to behave differently from regular days \citep{Krstonijevic2022adaptive}. Therefore, the same benchmark is run in the appendix  and shows that incorporating mobility data still significantly improves the forecasting performance when excluding holidays  (see Table \ref{table_score_target_agg}).
All these results confirm that adding mobility data leads to significant gains of approximately 10 $\%$ in forecasting the French electricity demand.

\subsection{Explaining the impact of mobility}
In this section we rely on variable selection to offer insight into the performance of the forecasts using mobility data. 
Moreover, we investigate the relation between the electricity demand and the work feature, being the second most explanatory variable.
\paragraph{Variable selection}
Combining the calendar, meteorological, electricity, and mobile network datasets results in 38 features. 
Moreover, some of these features are highly correlated, as explained in Appendix \ref{sec:region} for the \textit{temperature} and the \textit{school holidays} features.
Therefore, it is necessary to select a smaller number of features, with as uncorrelated effects as possible, in order to better understand how they relate to the electricity demand.
Nevertheless, the usual variable selection methods based on cross-validation \cite{wasserman2009high, Huang2010variable, marra2011practical} are not directly applicable to time series, because the samples are not independent.
To reduce  the dimension of the problem, one solution is to rank the features by order of importance \cite{genuer2010variable}.
In this paper, we consider three such methods: the mRMR ranking, the Hoeffding D-statistic ranking, and the Shapley  value ranking.
For multivariate time series, feature selection can be achieved thanks to the minimum redundancy maximum relevance (mRMR) algorithm, which consists in selecting variables that maximise the mutual information with the target \cite{peng2005feature, han2015joint}. 
As a result of running the \textit{mRMRe} package \cite{De2013rRMRe}, the most important variables, in decreasing order of importance, are the \textit{temperature}, the \textit{work} index, and the \textit{time of year}. 
The Hoeffding D-statistic ranking and the Shapley  value rankings  are detailed in Appendix \ref{sec:variable_selection}.
All three rankings are consistent, implying that the \textit{work} index is more important than the calendar data. 
As a result of this analysis, the \textit{tourism}  and \textit{residents} indices do not appear to have a significant effect on the French electricity demand.

\begin{figure}
    \centering
    \caption{\textbf{Effects of the features on electricity demand.}}
    \includegraphics[width = 0.49\textwidth]{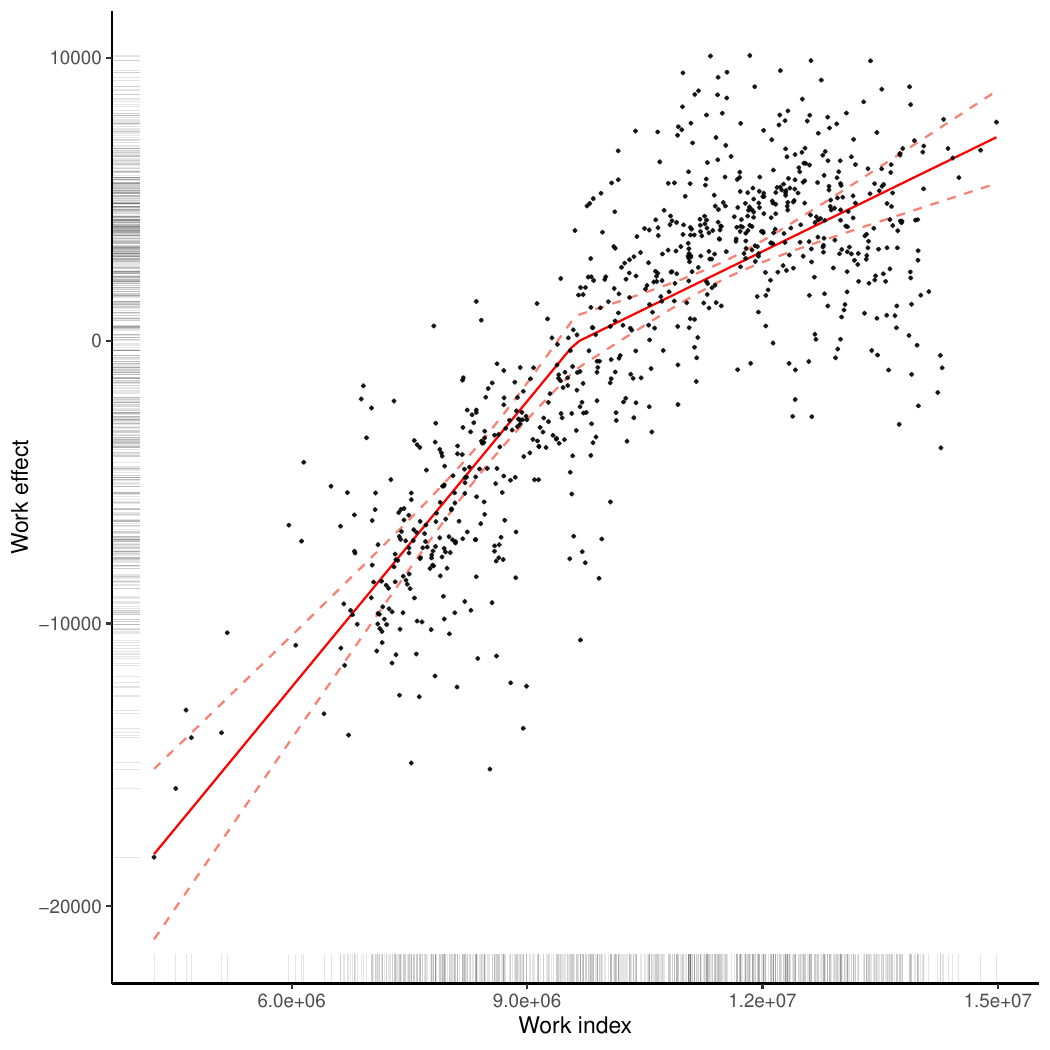}\includegraphics[width = 0.49\textwidth]{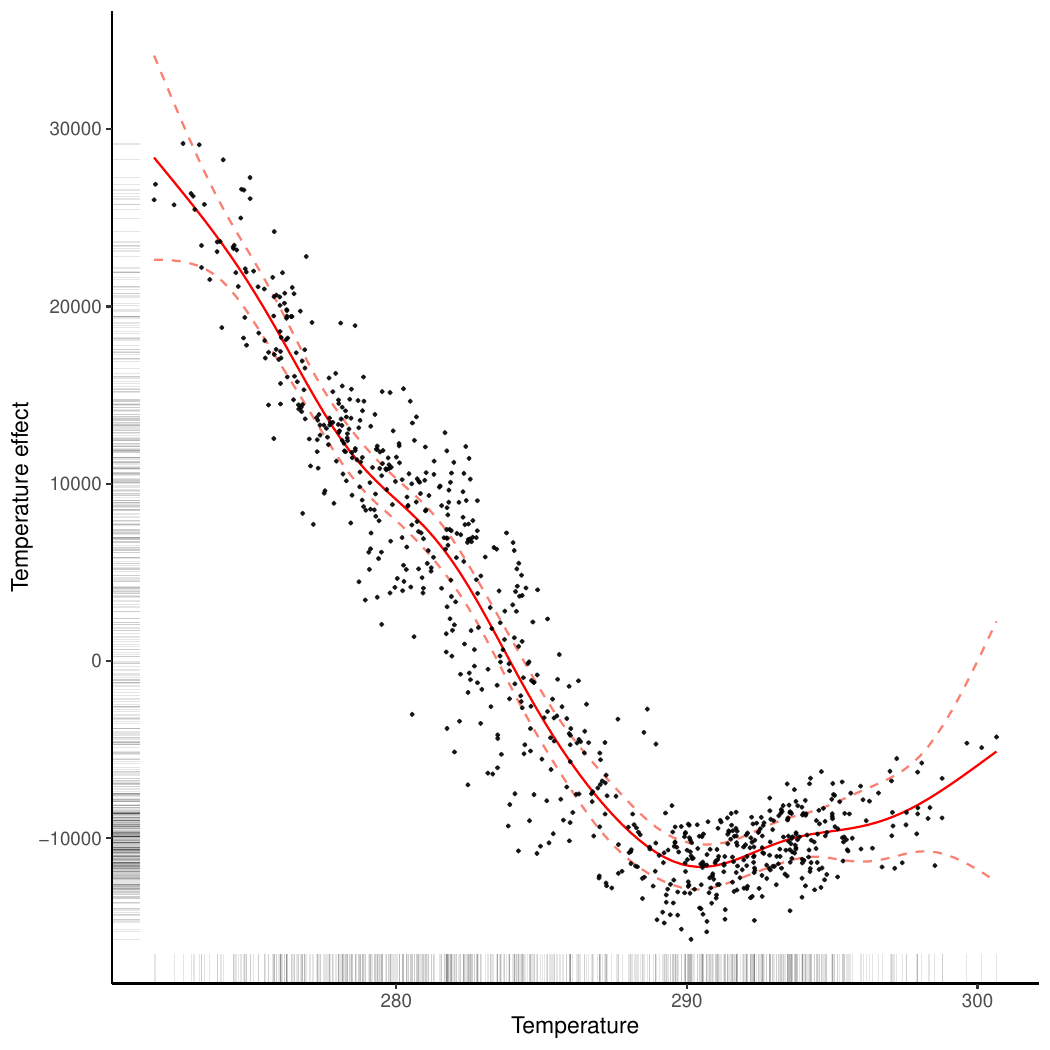}
    \flushleft
    {\small \textbf{Left:} Electricity load corrected from temperature as a function of work index. \textbf{Right:} Electricity load corrected from work as a function of work index. Each black point is an observation at 10 a.m.. The effect given by the GAM regression is shown in red.
    Dotted red lines corresponds to the 95$\%$ confidence interval of the effects.}
    \label{fig:gam_effect}
\end{figure}

\paragraph{Impact of work dynamics on the electricity demand}
\label{sec:stat_ana}
Outperforming the state-of-the-art indicates that the mobility data explain what happened during the sobriety period.
However, it does not provide any formal insights into the future performance of this index, which is why we run a statistical analysis of the predictive ability of mobility data.
Moreover, state-of-the-art data assimilation techniques being difficult to analyse, we restrain ourselves to statistical models of the electricity demand.
Since the effect of temperature is known to be non-linear, we consider GAMs instead of the usual linear regressions.
As suggested by variable selection methods, 
we consider the electricity demand corrected for the effect of temperature.
Figure  \ref{fig:gam_effect} shows that the electricity demand increases with the work index, i.e., the higher the number of people at work, the higher the electricity demand. 
\begin{figure}
    \centering
    \caption{\textbf{Dynamics captured by the work index.}}
    \includegraphics[width = 0.49\textwidth]{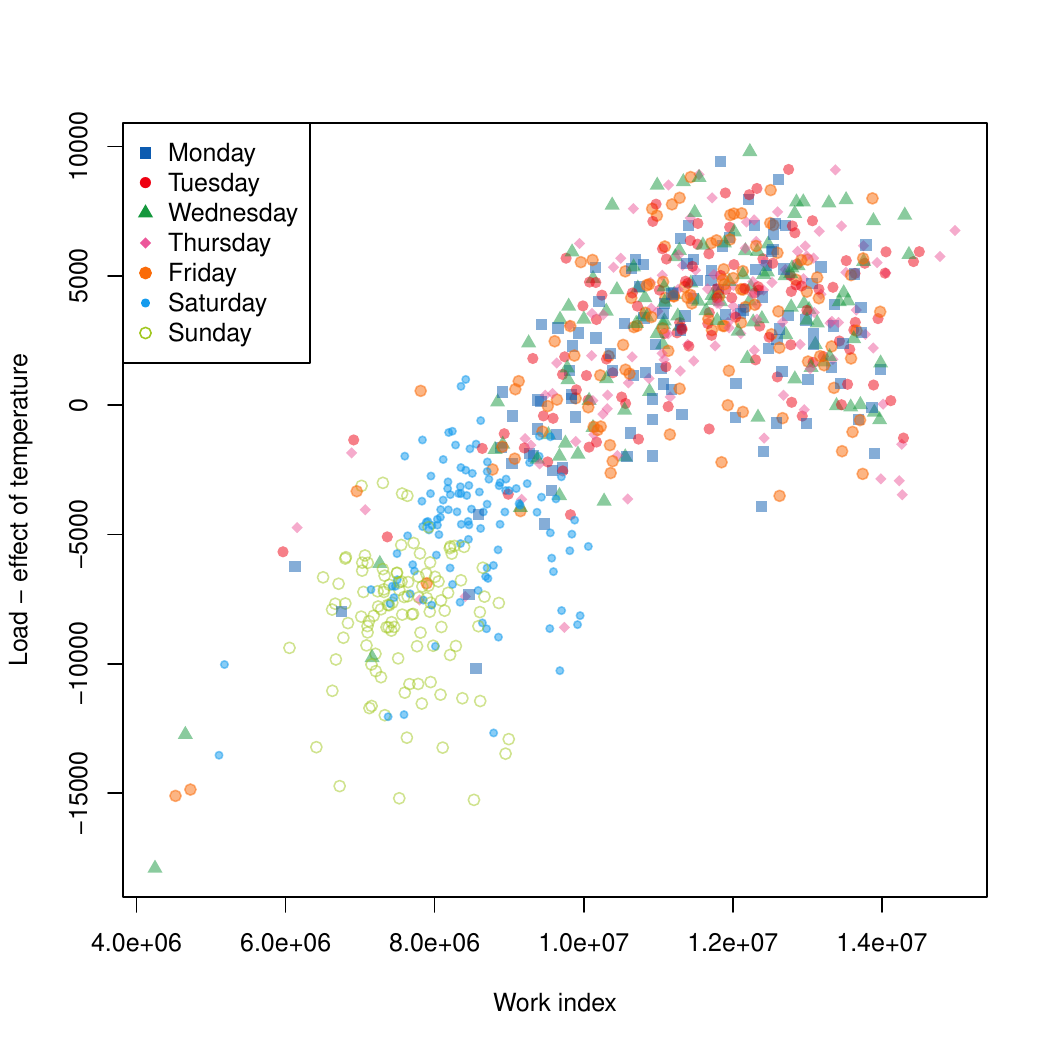}
    \includegraphics[width = 0.49\textwidth]{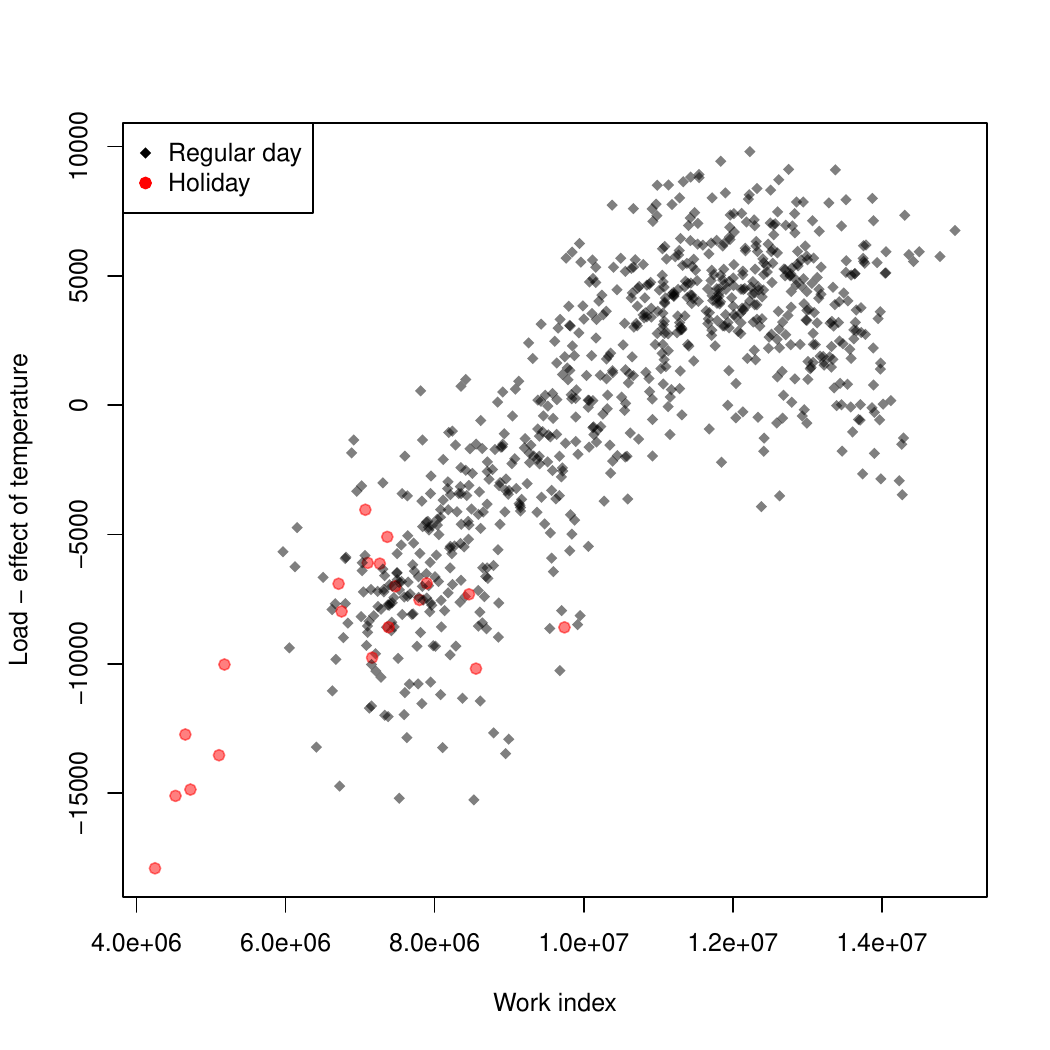}
    \flushleft
    {\small Each point is an observation of the residuals as a function of the work index at 10 a.m. between 07-2019 and 03-2022. \textbf{Left:} Dependence of the \textit{work} index on the day of the week. \textbf{Right:} Holiday pattern.}
    \label{fig:mob}
\end{figure}
Moreover, it manages to combine different dynamics.
First, Figure \ref{fig:mob} (left) shows how it accounts for the effect of weekends,  thus capturing the weekly seasonality related to work behaviour. 
Indeed, notice how working days, Saturdays (in purple), and Sundays (in yellow) correspond to different regimes of the work index. 
Figure \ref{fig:mob} (right) then shows that the work index explains consumption behaviour during the holidays (in red). 
Note that they have the same relationship as on regular days.
Moreover, the analysis of the impact of our \textit{work} index on the electricity demand, when fixing the day of the week and excluding holidays, shows that lower work dynamics correspond to a lower electricity demand (see Section \ref{sec:remote_wkg} in the appendix).
This shows that lower work  dynamics are associated with energy savings.
Further studies are needed to distinguish the role of remote working from the impact of other socio-economic parameters such as economic growth and  employment rate in this effect.
As expected, this effect is more pronounced during working hours.
As a result, the work index is more informative than calendar information alone. 
In fact, models using the \textit{work} index  perform better during the atypical event of the sobriety period than models based on calendar data, which only capture the seasonality in stationary signals (see Table \ref{table_GAM} in the appendix). 
This suggests that the \textit{work} index is explanatory of the electricity demand.

\section{Discussion}
In this work, we have shown that the period spanning from September 2022 to March 2023 was atypical in terms of the French electricity demand.
During this so-called sobriety period, we have observed a decrease in the electricity demand similar to what happened during the first COVID-19 lockdown.
However, this period of significant electricity savings lasted over six months, which is much longer than the COVID-related period of one month.
These observations are consistent with those of the French media and of the French transmission system operators. 
They prove that phenomena other than the annual seasonality and temperature are responsible for the recent significant changes in the electricity consumption behaviour.

To better understand  this collective energy-saving behaviour, we have introduced mobility indicators from mobile network data.
This is an original and efficient emerging way of tracking human mobility and assessing its impact on the electricity demand.
Indeed, the vibrancy of places varies over years, but also over the course of a day. These cycles are strongly linked to the people (both residents and non-residents) who visit them and to their behaviour. 
Some areas may be more attractive during the day or depending on seasons, while being quiet at night. This implies changes in the organisation and planning of such areas in terms of services, such as electricity demand, and more generally in terms of public policy.
For instance, following the COVID-19 lockdowns, remote working has been shown to have a significant impact on the electricity demand in the 2019-2020 period \cite{abu2020analysis, ku2022changes}.
Understanding the spatial practices of human populations is therefore fundamental to operational decisions and research in many fields. Individuals circulate in many places throughout the day - on average between 2.5 and 4 per person in French metropolitan areas \citep{mtect} - month, and year, whether for housing, work, education, personal relationships, or leisure.
For a long time, the main source of this information has been population censuses, supplemented by daily and tourist travel surveys, though they suffer from several limitations, including infrequent sampling, lack of reliability, and cost. 
One response has been the processing of digital traces, in particular data from mobile networks. 
The quality of these data has evolved with both the development of network technology and the ability to account for intrinsic biases, such as representation and precision.
Our high quality dataset was designed to precisely quantify human presence over France, at a very high frequency as compared to census, and has already been studied as such to account for residential behaviour \citep{Levy2023who}. 
Its advantage is that it allows not only to quantify with a high degree of accuracy the population present at a given time and place, but also to characterise the way they inhabit it (i.e., residing, working, exploring, or crossing).
Here, we have shown that these indices are highly correlated with other public datasets, and that they also account for human presence dynamics related to tourism and work. 
Such indices could become particularly relevant to track the rapid changes in work organisation catalysed by the COVID-19 health crisis, which have the potential to radically shape many aspects of societies and economies \cite{vyas2022new}.  
They also offer a great flexibility in terms of the scale (national, regional, city...) at which we study mobility \cite{pora2023telework}. 
In fact, in France, the collection of statistical information on remote working has traditionally been limited to ad hoc flash surveys. It is only from 2021 that questions on remote working will be included in the Continuous Employment Survey (the main reference statistical survey on the labour market) and from 2022 in the census, but with much less frequent sampling compared to mobile phone data \cite{pora2023telework}.

As evidenced by our benchmark in Table \ref{table_score_target_agg}, standard statistical models such as GAM struggled during the sobriety period. Indeed, for the same state-of-the-art GAM, the RMSE and the MAPE when excluding holidays are respectively $55 \%$ and $87 \%$ higher than in the same test period two years earlier (i.e., from September 2019 to March 2020) \cite{obst2021adaptative}. 
However, relying on a benchmark of statistical models, data assimilation techniques, and machine learning architectures, we have shown that using these data improves the state-of-the-art performance by an order of magnitude of 10\% during the sobriety period. 
Although evaluating the cost of load forecasting error is a difficult task, it has been estimated that a 1 $\%$ reduction in load forecasting error would save an energy provider up to an order of magnitude of one hundred thousand dollars per year per GW peak \cite{hong2016probabilistic}.  
Therefore, the gain of 0.2 $\%$ of MAPE resulting from exploiting mobility data is very promising.

In addition, we have shown that the \textit{work} index accounts for several dynamics including the impact of weekends and holidays on the electricity demand.
Notice how these dynamics are not specific to the sobriety period, which suggests that the benefits of using mobility data would generalise to the post-crisis period.
Overall, the higher the \textit{work} index, the higher the electricity demand. 
Future lines of research include studying the work index at a 1-hour frequency, over longer periods, and at the finer geographical scale of the French administrative regions.
Indeed, as shown in the Appendix \ref{sec:region}, mobile network data effectively capture human spatial dynamics other than those related to work, such as residence and tourism. Although in this paper we have not found a significant effect of such dynamics on national electricity demand, they might become visible when working at the regional level. 
Although we have shown that a reduction in the \textit{work} index corresponds to a reduction in the electricity demand, further studies are needed to disentangle the effect of economic growth, the employment rate, and remote working in this phenomenon.
Moreover, we have focused this work on mean forecast performance, i.e., on the ability of the forecast to predict the mean value of the electricity demand.
Another interesting topic would be  to evaluate the variance of the electricity demand given the \textit{work} index, which is useful for practitioners when acting on the electricity market.
Finally, in practice, it would take a few days to clean, aggregate and adjust the indices. 
For operational use, further studies are needed to quantify the impact of a delay in the use of the \textit{work} index on the performance of benchmark forecasts, or conversely, to study the predictability of the \textit{work} index.

\section{Methods}
\subsection{Open calendar, meteorological, and electricity datasets}
The reference dataset runs from 2013-01-08 to 2023-02-28. It consists of calendar data (dates and holidays), meteorological data (temperature), and historical data (electricity power load at different scales). 
All these data are public and distributed under the Etalab open source licence. 
The calendar data are extracted from the French open source database \citep{dataGouvJourFeries, dataGouvVacances}. 
It regroups the holiday periods according to the three French holiday groups --- as in France, the holidays depend on the region you live in---, as well as the French national holidays.
As the holidays are well known French conventions, this calendar dataset has no missing values. 
The meteorological data are extracted  from the SYNOP Météo-France database \citep{meteoFrance}.
Météo-France is the French public agency responsible for the national weather and climate service. 
The dataset consists of 3-hourly temperature measurements from 62 meteorological stations located throughout the French territory.
This dataset has many missing values, which are filled as follows. First, if a station has a missing value at time $t$ and the station's measurements are available 3 hours before and 3 hours after $t$, the missing value is filled with the mean of these two measurements. Then, if no such values are available, the missing temperature is approximated by the temperature of the nearest station. Finally, if all stations in a region have missing values, their temperature is estimated by taking the mean of the temperature at the same hour the day before and the day after.  
Finally, the historical electricity load dataset is extracted from the RTE public releases \citep{rteData}.
RTE (Réseau de Transport d'Electricité) is a the French Transmission System Operator.
It provides high quality data on regional electricity consumption in France, with a frequency of 30 minutes. The national electricity load has no missing values, which is valuable since this is the final target all along this article.

\subsection{Mobility dataset}
\label{sec:mob_dataset}
The reference dataset is  complemented by mobility indices. 
These mobile phone data were provided by the Flux Vision  business service of Orange \cite{fluxVision}, in the form of presence data reports. 
These include the number of visitors in 101 geographical areas of mainland France, which represent the second level of national administrative divisions. 
For each location and each day, the data are stratified by the type of visitor (resident, usually present, tourist, excursionist, recurrent excursionist) and origin (foreign, local, non-local). 
Mobile phone data have been previously anonymised in compliance with strict privacy requirements and audited by the French data protection authority (Commission Nationale de l'Informatique et des Libertés). 
The computation of the presence data reports is based on the on-the-fly processing of signalling messages exchanged between mobile phones and the mobile network, usually collected by mobile network operators to monitor and optimise the mobile network activity. 
Such messages contain information about the identifiers of the mobile subscriber and of the antenna handling the communication, the timestamp and the type of event to be recorded (e.g., voice call, SMS, handover, data connection, location update). 
Knowing the location of the antennas makes it possible to reconstruct the
approximate position of the device in communication. This was then used to compute
the total number of visitors, with no residual information tracing back to the
individual users. 
Each visitor was then characterised based on the time spent and their origin. 
More specifically: 
\begin{itemize}
\item Resident: person whose main  area of attendance is in the study area and who has spent at least 22 nights (not necessarily consecutive) there.
\item Usually present: person who is not a resident of the study area but has been seen in the study area repeatedly: more than 4 times in different weeks in the last 8 weeks.
\item Tourist: person spending the night in the study area who is neither resident nor usually present.
\item Excursionist: person not staying overnight the night before and the night of the study day, and present less than 5 times during the day in the last 15 days.
\item Recurrent excursionist: person who has not spent the night before and the current night in the study area and who has been present more than 5 times during the day in the last 15 days.
\end{itemize}
Moreover, their origin is categorised as follows:
\begin{itemize}
\item Foreign: person with a foreign SIM card.
\item Local: person with a billing address in the study area.
\item Non-local: person with a billing address outside the study area.
\end{itemize}
This data is then adjusted by Orange Flux Vision to account for spatial and temporal biases and to be representative of the general population. 
To do so, they use spatially stratified market share data, socio-economic data from the national statistics institute Insee (Institut national de la statistique et des etudes economiques), mobile phone ownership data from Insee, and
customer socio-demographic information provided upon subscription.
From these data, we construct three indices. The \textit{work} index is the sum of the the recurrent excursionists. The \textit{tourism} index is  the sum of foreign and non-local tourists. The \textit{resident} index is  the sum of all the residents and usually presents.
In this article, the mobility dataset covers the periods from 2019-07-01  to 2020-03-01, from 2020-07-01 to 2021-03-01, from 2021-07-01 to 2022-03-01, and from 2022-07-01 to 2023-03-01.

\subsection{Benchmark models}

Models are then evaluated according to the following test errors. Let $T_{test}$ be the test period, $(y_t)_{t \in T_{test}}$ be the target, and  $(\hat y_t)_{t \in T_{test}}$ be an estimator of $y$. The root mean square error is defined by
$\mathrm{RMSE}(y, \hat y) = (\frac{1}{T_{test}}\sum_{t= \in T_{test}} (y_t-\hat y_t)^2)^{1/2}$
and the mean absolute percentage error is defined by
$\mathrm{MAPE}(y, \hat y) = \frac{1}{T_{test}}\sum_{t \in T_{test}} \frac{|y_t-\hat y_t|}{|y_t|}$.
Both these errors are useful for operational uses. Since samplings of time series are dependent, confidence intervals are obtained by bootstrapping \cite{lahari2023resampling}. All benchmark models are direct adaptations of state-of-the-art models in the French electricity demand forecasting.
The GAM is extracted from \cite{obst2021adaptative}. 
The static and dynamic Kalman filters are inspired from \cite{vilmarest2022state}. 
The Viking algorithm comes from \cite{vilmarest2023adaptative}.
The GAM boosting parameters are from \cite{bentaieb2014a}.
The random forest and random forest with bootstrap parameters are taken from \cite{gohery2023random}.
A full description of the models can be found in the appendices.

\section{Data and code availability}

The source code detailing how to create and update the electricity dataset and all the models used in this article are available at \href{https://github.com/NathanDoumeche/Mobility_data_assimilation}{https://github.com/NathanDoumeche/Mobility\_data\_assimilation}.
Therefore, the change point method of Figure \ref{fig:obst}, as well as the dataset and the benchmarks without mobility data of Table \ref{table_score_target_agg2} are directly reproducible for future research and can be updated to different periods of interest.
However, mobility indices are not openly available.

\bmhead{Acknowledgments}

This work was funded by Sorbonne University, EDF R\&D, and Orange Innovation Research. The authors bear full responsibility for the conclusions and findings.

\bibliography{biblio}

\pagebreak
\begin{appendices}

\section{Datasets and features}
\label{sec:data}
In this appendix, we provide further insights into the exploratory analysis of the mobility dataset. We show how the indices also capture holiday dynamics at the regional level, by comparing the mobile network-based \textit{tourism} index with official tourism statistics from Insee, and by studying the temporal evolution of the \textit{work} index.

\subsection{Regional human presence indices}
\label{sec:region}
\begin{figure}
    \centering
    \caption{\textbf{Regional indices.}}
    \includegraphics[width =0.9 \textwidth]{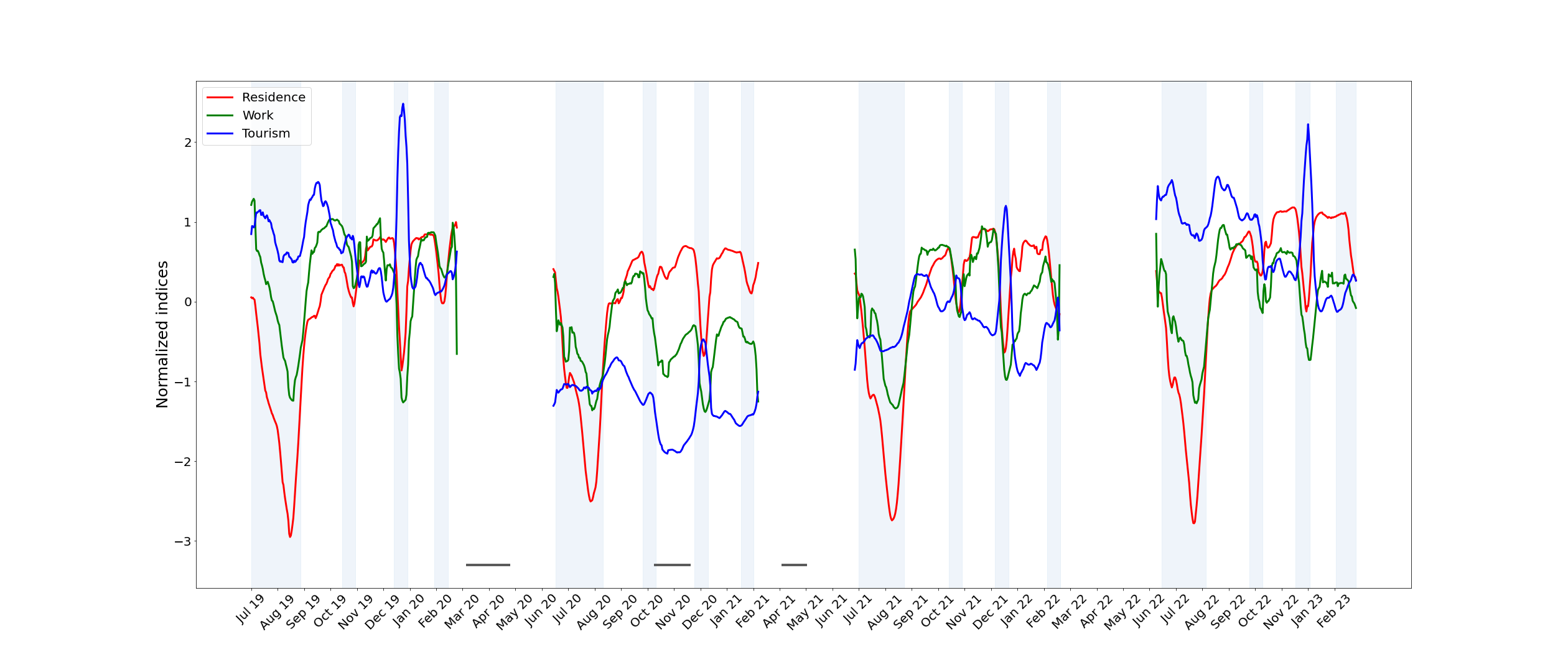}
    \includegraphics[width = 0.9\textwidth]{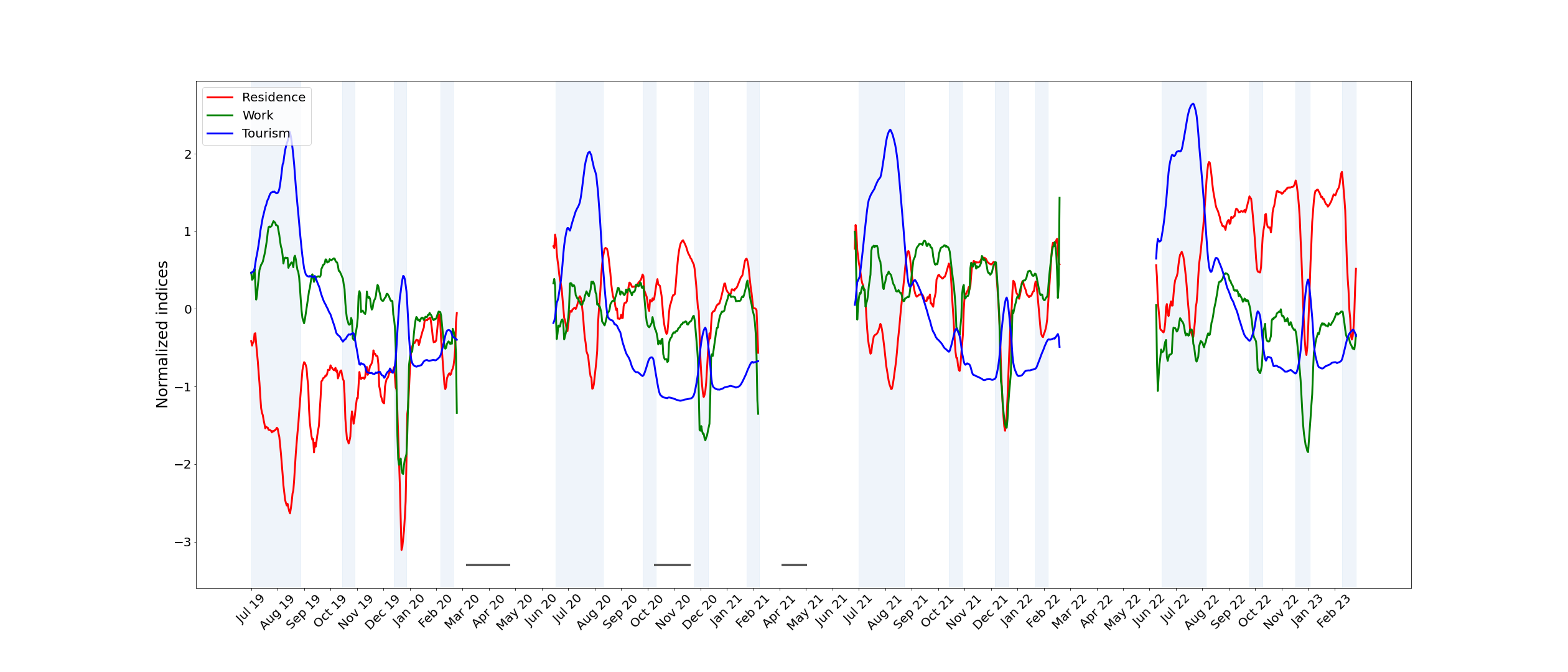}
    \flushleft
    {\small 7-day rolling average of the mobility indices for the IDF (top) and PACA (bottom) regions. Indices have been normalised, i.e., subtracted their empirical mean and divided by their empirical standard deviation. The mobile network dataset only covers the period from July to March each year.
    Coloured areas correspond to the region school holidays, and horizontal grey lines mark the three main COVID-19 lockdowns in France}
    \label{fig:IDF_timeline}
\end{figure}

Although for the purpose of the national forecasts we have only relied on national level indices, mobile network data were also available at a regional level, which helps to better understand the data at hand.
In order to obtain a preliminary understanding of the data, we compute the Pearson product-moment correlation coefficient $r$ between the human presence variables on the one hand and the calendar and meteorological data on the other, for the different regions of mainland France. 
This analysis confirms that our indices follow several well-known human spatial dynamics.
In large urban regions, such as the Île-de-France (IDF), we observe negative correlation between the \textit{residence} index and both the calendar variables \textit{school holidays}  and \textit{summer holidays} ($r =-0.65$ and $r =- 0.84$, respectively), as well as \textit{temperature} ($r =-0.70$).
This captures how IDF residents leave their region during the holidays and then behave as tourists.
Consistently, in regions that are traditionally popular holiday destinations, such as the coastal region of Provence-Alpes-Côte d'Azur (PACA), the \textit{tourism index} variable is positively correlated with the calendar variables \textit{school holidays}  and \textit{summer holidays} ($r =0.58$ and $r =0.86$, respectively) and the meteorological variable \textit{temperature} ($r =0.82$). 
The \textit{work} index  shows a similar behaviour in both categories of regions, with a negative correlation with the \textit{weekly holiday} calendar variable ($r = - 0.54$ in IDF and $r = - 0.55$ in PACA). 
Seasonal changes in the distribution of the different categories of population are more evident when looking at the timeline. 
To better characterise these patterns according to the specificity of the territories, we show the evolution of the daily \textit{tourism}, \textit{residence}, and \textit{work} indices in IDF (Figure \ref{fig:IDF_timeline} - Top) and PACA (Figure \ref{fig:IDF_timeline} - Bottom). In line with the Pearson correlation, in a region with a high level of economic activity such as IDF, the \textit{residence} and \textit{work} indices tend to increase during the off-peak periods and to decrease during the holidays. We observe the opposite behaviour for the \textit{tourism} index in PACA, which is a very touristic region. Moreover, unlike in IDF, the \textit{work} index in PACA does not decrease significantly during the summer holidays.  
This could be explained by the different composition of the labour markets, with a high proportion of workers in tourism in PACA.

We can also clearly see the effects of the COVID-19 health crisis. In IDF, for example, the \textit{tourism} and \textit{work} indices significantly dropped during the crisis. 
They then gradually increased in the post-COVID period, but without reaching the pre-COVID levels.
This is especially pronounced for the \textit{work} index, probably because of the changes in work organisation triggered by the health crisis and also as an effect of the energy crisis. In PACA, on the other hand, we observe a less important impact on tourism, partly due to a different seasonality and origin (there are more local tourists, i.e. who do cross the borders) than in IDF. Interestingly, the \textit{residence} index seems to be gradually increasing since the COVID-19 crisis. This phenomenon of migration to certain regions of France has been documented by Insee in the report \cite{michailesco2023migrations}, but would deserve a more in-depth analysis.

\subsection{Tourism index from mobile-phone data}
\begin{figure}
    \centering
    \caption{ \textbf{Comparison of the Insee and the mobile network tourism indices.}}
    \includegraphics[width = 0.49\textwidth]{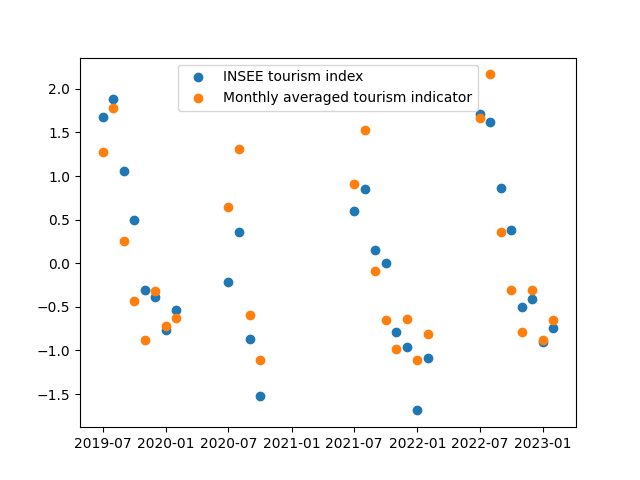}
    \label{fig:tourism}
\end{figure}

Evaluations of the number of tourists and residents has been shown to be slightly correlated with electricity demand in highly touristic areas \cite{Bakhat2011estimation, lai2011the}. 
This is why we have created and studied a \textit{tourism} index at the national level.
Traditionally, most of these assessments have been carried out on an annual or monthly basis. 
A strength of our mobile phone-based \textit{tourism} index is that it can be calculated at finer temporal and geographical scales. 
To further assess its performance as a proxy for tourism activity, we compare its monthly
average with the Insee tourism index \cite{INSEEData}, as shown in Figure \ref{fig:tourism}.
As a result, we obtain an 87 \% correlation between the two signals, showing that the tourism index efficiently captures tourism trends. However, our study found that tourism has no significant impact on French electricity demand (see Appendix \ref{sec:variable_selection}).

\subsection{Work index and calendar features}
\begin{figure}
    \centering
    \caption{\textbf{Residuals as a function of the work index over the years.}}
    \includegraphics[width = 0.49\textwidth]{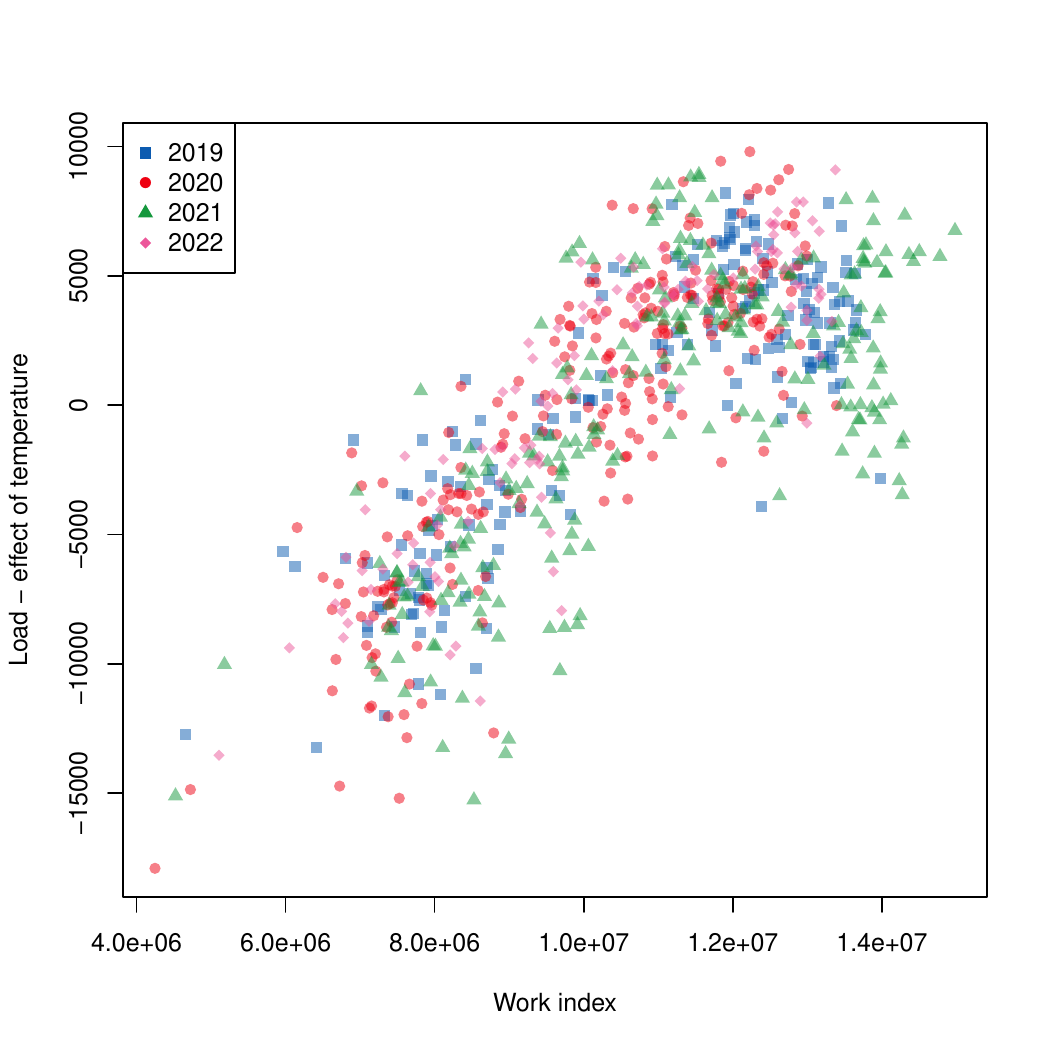}
     \flushleft
     {\small Each point is an observation between 07-2019 and 03-2022.}
    \label{fig:year}
\end{figure}

\begin{figure}
    \centering
    \caption{\textbf{Comparison of work indices.}}
    \includegraphics[width = 0.49\textwidth]{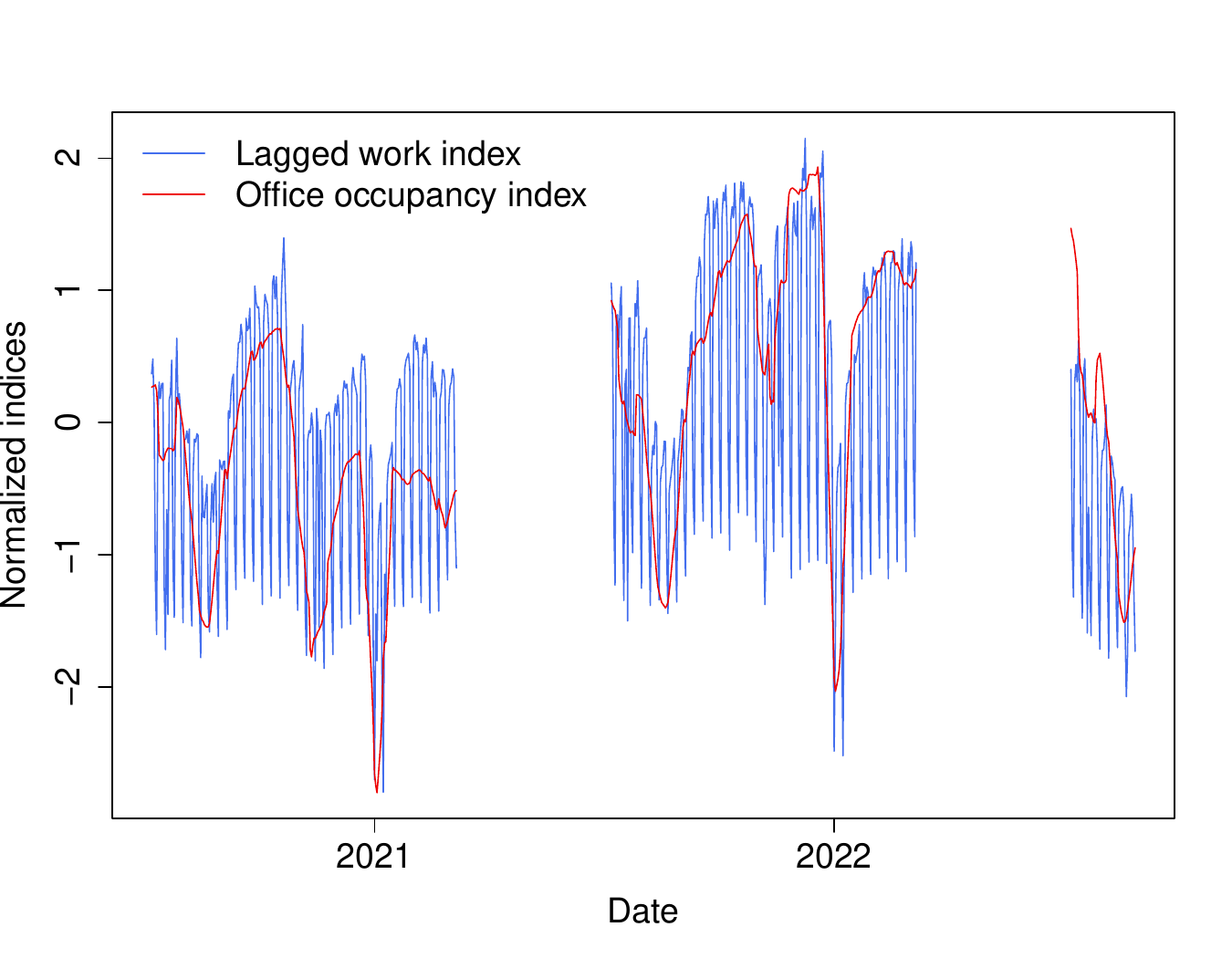}
    \includegraphics[width = 0.49\textwidth]{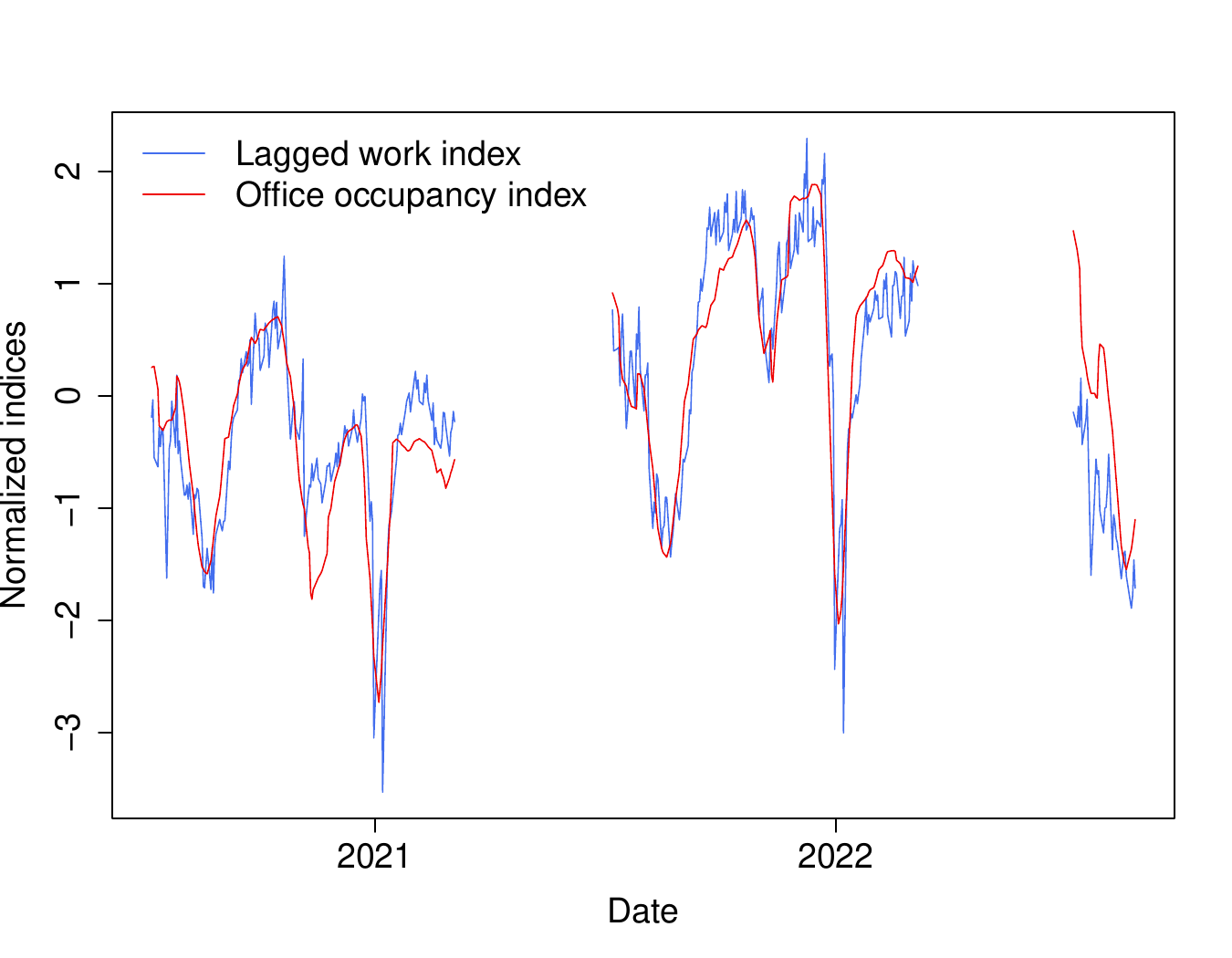}
    \flushleft
    {\small Comparison of the 7-day lagged mobile phone based index and the normalcy office occupancy index on all days (left), and 
    when excluding weekends and the holidays (right).}
    \label{fig:normalcy2}
\end{figure}

\begin{figure}
    \centering
    \caption{\textbf{Electricity demand corrected for temperature as a function of the work index over the day.}}
    \includegraphics[width = 0.49\textwidth]{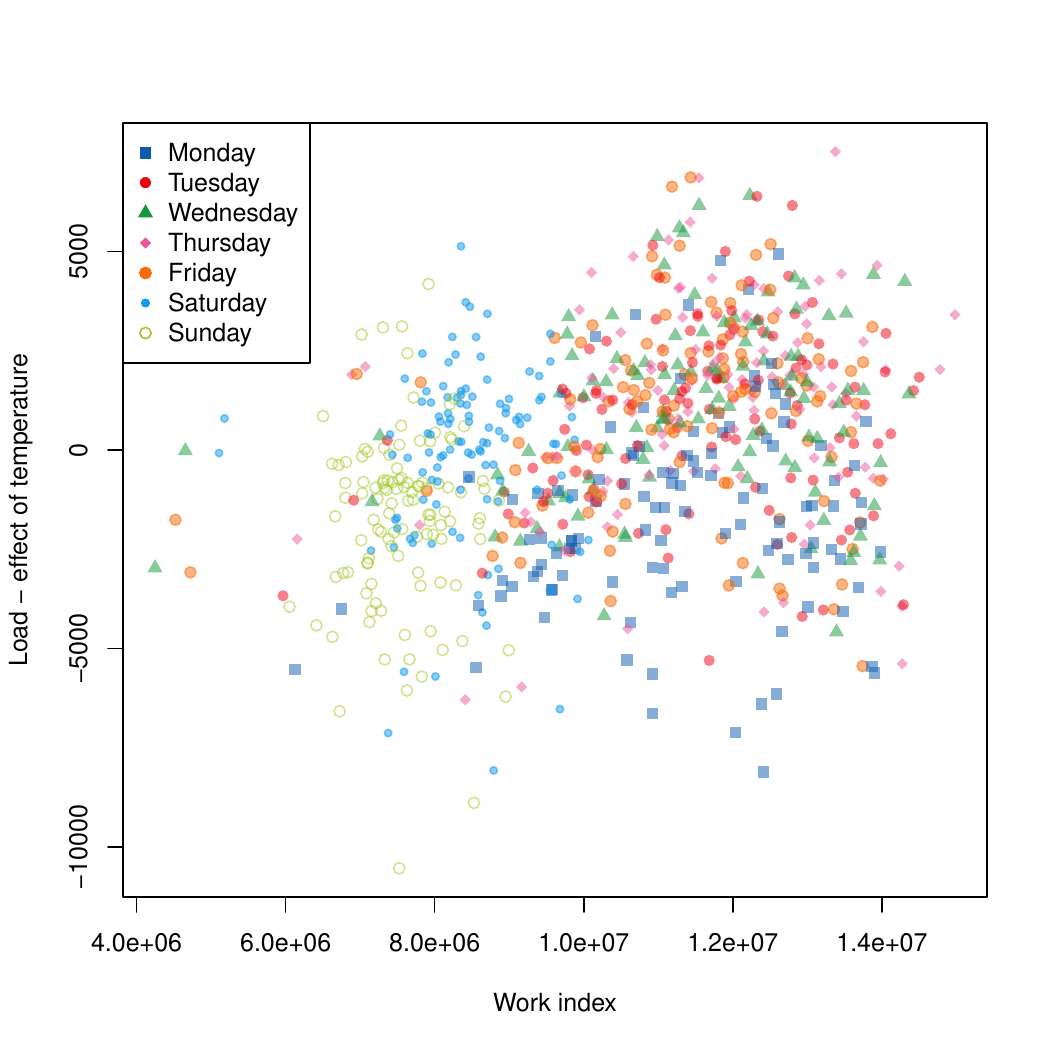}
    \includegraphics[width = 0.49\textwidth]{day_dependency.pdf}
    \flushleft
    {\small Each point is an observation between 07-2019 and 03-2022. \textbf{Left:} 2 a.m. \textbf{Right:} 10 a.m.}
    \label{fig:hours_comparison}
\end{figure}

As explained in Introduction, several phenomena occurred between 2020 and 2023 that significantly changed human behaviour and affected the French electricity demand. 
To better understand the impact of the \textit{work} index on the electricity demand, it is therefore important to see whether this dependence has changed over time.
In fact, Figure \ref{fig:year} shows that the dependence of electricity demand in the \textit{work} index has been stationary over the years.
This shows that this relationship has been robust to the aforementioned events, which is an argument to believe that the results of this article will generalise well to future periods of interest.
In addition, as shown in Figure \ref{fig:normalcy2}, unlike The Economist's office occupancy index, our \textit{work} index from mobile data captures the reduction in work activity due to weekends and holidays. 
Furthermore, as expected, Figure \ref{fig:hours_comparison} shows that the \textit{work} index is only useful for electricity demand forecasting during working hours.
Indeed, the electricity demand corrected for the temperature effect has a clear dependence in the index at 10 a.m., but not at 2 a.m..
This will be studied in more details in Appendix \ref{sec:remote_wkg}.

\section{Benchmark and models}
\label{app:Benchmark}
In this appendix, we detail the framework and the models of Table \ref{table_score_target_agg2}.
\subsection{Handling missing values in mobile network data}
\label{app:missing}
There are two types of missing data in our datasets. To begin with, the datasets are regularly sampled time series with different frequencies. Indeed, recall that the calendar and the electricity datasets have a 30-minute frequency, while the meteorological dataset has a 3-hour frequency, and the mobile phone dataset has a 1-day frequency.
A common method to deal with differences in sampling frequency is to impute the missing value by interpolation \cite{emmanuel2021a}.
The interpolation method for meteorological data is described in the Methods section of the main paper, while while the Orange indices are kept constant throughout the day.

Moreover, the mobile network dataset only covers the periods from 2019-07-01  to 2020-03-01, from 2020-07-01 to 2021-03-01, from 2021-07-01 to 2022-03-01, and from 2022-07-01 to 2023-03-01. Though various techniques have been developed to tackle samplings irregularities in time series \citep[see, e.g.,][]{shukla2021survey}, dealing with large sets of consecutive missing values is still very challenging.
The three main approaches when studying time series with consecutive missing values are  deletion, imputation, and imputation with masks \cite{emmanuel2021a}.
Deletion consists of discarding any observation with at least one missing value.
Though this is the simplest way to deal with missing values, it can  introduce a bias if the missing data are not-at-random, i.e., if the missing data are actually informative about the target \cite{little2019statistical}. 
In a regression task, imputation techniques aim to fill in the missing values.
The state-of-the-art in time series imputation is very broad, and this is an active field of research \citep[see, e.g.,][]{ma2020transfer}.
However, the imputation that maximises the regression models performance is not necessarily the one that reconstructs the missing values most accurately  \cite{zhang2021missing, ayme2023naive}.
This makes it more difficult to understand and explain the real effect of the imputed features on the target variable.
Imputation with mask consists of imputing the missing values and keeping track of which observations have been imputed by adding a new feature equal to 1 if the observation comes from an actual measurement and 0 if it was imputed.
In this paper, the pattern of missing data is regular, spanning each year from March to July, and does not depend on the  explanatory variables (temperature, \textit{work} index...).
Thus, to simplify the analysis, we have chosen the deletion framework and have not tried to impute the missing values of the mobile network indices.

In fact, the aim of Table \ref{table_score_target_agg2} is not only to show that mobile phone indices help to improve the performance of state-of-the-art forecasting algorithms, but also to attest that this is still true even when comparing the complete open dataset with the incomplete mobile phone dataset.
Indeed, on the one hand, models "without mobility data" are trained on the complete open calendar, weather, and electricity datasets spanning from 2013-01-08 to 2022-01-09.
On the other hand, models "with mobility data" are created in two steps, according to the transfer learning framework presented in \cite{antoniadis2021hierarchical}.
First, a model trained without mobility data from 2013-01-08 to 2022-01-09 provides an estimate $\hat{Load}$ of the electricity demand $Load$.
Then, another model is trained in the deletion framework to forecast the error $err = Load - \hat{Load}$, also called the residual, using the mobile phone dataset.
This second forecast is denoted by $\hat{err}$.
The final forecast is therefore the sum of the two forecasts $\hat{Load} + \hat{err}$.
Notice that, this framework advantages the reference forecast "without mobility data".
In fact, the gains from using mobile phone data are much higher if the training periods of all models are restricted to the period for which mobile phone data are available (although we have not included these results in the paper for the sake of simplicity).
However, the framework we have chosen allows us to assess the interest of using mobile phone data from an operational point of view.
It ensures that the best models trained using the mobile phone dataset outperform the best models trained on the full open datasets. 
Therefore, the gains of $10 \%$ we have obtained should be much higher if we had access to a more complete mobile phone dataset.
We have chosen the residual method to account for the mobile phone data because it gives better results than directly training models "with mobility data" on all datasets restricted  to the period for which the mobile phone data are available (once again, we have not included these results in the paper for simplicity).

\subsection{Statistical models}

\subsubsection{Time series models}
Persistence models are the simplest models for time series. They consist of estimating the target with its own lags. They are common baselines in time series benchmarks because of their simplicity, their ability to capture trends, their explainability, and their robustness to ruptures in the data distribution. In Table \ref{table_score_target_agg2}, the persistence estimator is the 24-hour lag of the electricity demand.

Seasonal Autoregressive Integrated Moving Average (SARIMA) models \cite{boxjen76} are very common for time series analysis. Here, we train one model for each of the 48 half-hours in a day to capture the daily seasonality of the data. 
Each model is then fitted with a weekly seasonality by running the \texttt{auto.arima} method of the \texttt{forecast} package in \texttt{R}.

\subsubsection{Generalized Additive Models}
Generalized Additive Models (GAMs) are a generalisation of linear regression. 
Instead of learning linear coefficients linking some features $\boldsymbol{x} = (x_{1},...,x_{d})$ to a target $y$, a GAM learns the nodes and the coefficients of the regression of the features on the targets with respect to a basis of splines. 
More precisely, given a target time series $y = (y_t)_{t \in T}$ on a time horizon $T$, and some time series explanatory variable $\boldsymbol{x} = (x_{t,1},...,x_{t,d})_{t \in T}$, the response variable $y$ is decomposed as
\begin{equation*}
    y_t = \beta_0 + \sum_{j=1}^d f_j(x_{t,j}) + \varepsilon_t \,,
\end{equation*}
where $\varepsilon = (\varepsilon_t)_{t\in T}$ are independent identically distributed (i.i.d.) random noises.
Though the target $y_t$ at time $t$ is a real number, each explanatory time series $x_{k} = (x_{t,k})_{t \in T}$ at time $t$ has a dimension $d_k \geqslant 1$.
Therefore, non-linear effect of multiple variables are allowed, such as $y_t = \beta_0 + f_1(x_{t, 1}, x_{t_2}) + \varepsilon_t$.
The goal of GAM optimisation is to find the best non-linear functions $f_1, \hdots, f_d$ to fit $y$.
Thus, each non-linear effect $f_j$ is decomposed on a spline basis $(B_{j,k})_{1\leqslant j \leqslant d,\; k \in \mathbb{N}}$ with coefficients $\boldsymbol\beta_j$ such that
\begin{equation*}
    f_j(x) = \sum\limits_{k=1}^{m_j} \beta_{j,k} B_{j,k}(x) \,.
\end{equation*}
where $m_j$ corresponds on the dimension of the spline basis. 
The functions $f_j$'s are centred. The coefficients $\beta_0, \boldsymbol{\beta}_1, \dots, \boldsymbol{\beta}_{d}$ are obtained by penalised least squares. The penalty term involves the second derivatives of the functions $f_j$, forcing the effects to be smooth (see \cite{wood2017generalized}).

The GAM model used in our experiments presented in Table   \ref{table_score_target_agg}  is taken from \cite{obst2021adaptative}.  
As it is usual in load forecasting with GAMs we consider one model per half-hour of the day, with the 48 half-hour time series considered independently.
Therefore, 48 models are fitted, one for each half-hour of the day.
Given a half-hour $h$, our model is
\begin{align}\label{eq:modele_GAM_FR}
    \text{Load}_{h,t} =\ & \sum_{i=1}^7 \sum_{j=0}^1 \alpha_{h,i,j}\;\boldsymbol{1}_{\text{DayType}_t=i}\;\boldsymbol{1}_{\text{DLS}_t=j} \nonumber\\
    & + \sum_{i=1}^7 \beta_{h,i}\; \text{Load1D}_t \;\boldsymbol{1}_{\text{DayType}_t=i} + \gamma\; \text{Load1W}_t  \\
    & + f_{h,1}(t) + f_{h,2}(\text{ToY}_t) + f_{h,3}(t,\; \text{Temp}_{h,t}) + f_{h,4}(\text{Temp95}_{h,t}) \nonumber\\
    & + f_{h,5}(\text{Temp99}_{h,t}) + f_{h,6}(\text{TempMin99}_{h,t},\; \text{TempMax99}_{h,t})  + \varepsilon_{h,t} \nonumber\,,
\end{align}
where the timestamp $t$ is the day, and 
\begin{itemize}
    \item
    $\text{Load}_{h,t}$ is the electricity load on day $t$ at instant $h$.
    \item
    $\text{DayType}_t$ is a categorical variable indicating the type of the day of the week.
    \item
    $\text{DLS}_t$ is a binary variable indicating whether $t$  is daylight saving time or standard time.
    \item
    $\text{Load1D}$ and $\text{Load1W}$ are the load of the previous day and the load of the previous week respectively.
    \item
    $\text{ToY}_t$ is the time of year, growing linearly from 0 on the 1\textsuperscript{st} of January 00h00 to 1 on the 31\textsuperscript{st} of December 23h30.
    \item
    $\text{Temp}_{h,t}$ is the national average temperature at time $h$ on day $t$.
    \item
    $\text{Temp95}_{h,t}$ and $\text{Temp99}_{h,t}$ are exponentially smoothed temperatures of factor respectively $\alpha=0.95$ and $0.99$. For example,  $\alpha=0.95$ corresponds to
    \[\text{Temp95}_{h,t}=\alpha \text{Temp95}_{h-1,t} + (1-\alpha) \text{Temp}_{h,t}.\]
    \item
    $\text{TempMin99}_{h,t}$ and $\text{TempMax99}_{h,t}$ are respectively the minimal and maximal value of $\text{Temp99}$ on day $t$ on all instants $i$ such that $i \leqslant h$.
\end{itemize}
These models are then implemented in \texttt{R} by using the \texttt{mgcv} library \cite{wood2015package}. 
We have used the default thin-plate spline basis to represent the $f_j$'s, except for the time of year effect $f_2$ for which we choose cyclic cubic splines (see \cite{wood2017generalized} for a full description of the spline basis). 
The dimensions of the bases are usually less than $5$, except for $f_2$ which has a basis of dimension $20$.

\subsection{Data assimilation techniques}

\subsubsection{State Space Models}
State space models are efficient to capture time-varying structures (as opposed to seasonality) in time series \cite{hyndman2008forecasting}.
In particular, the Kalman filter is a powerful mathematical and algorithmic tool introduced by \cite{Kalman1960} for state space model estimation. 
In electricity load forecasting, Kalman filters are used to adjust the output of a GAM using recent observations of electricity demand \cite{vilmarest2022state}. 

Following the notation of Equation \eqref{eq:modele_GAM_FR}, let $f(\boldsymbol{x}_t) = (1,\overline{f}_1(x_{t,1}),...,\overline{f}_d(x_{t,d}))^\top$ where $\overline{f}_j$ is the normalisation of $f_j$. 
Our goal is to estimate a time-varying vector $\boldsymbol\theta_t \in \mathbb{R}^{d+1}$ such that $\mathbb{E}[y_t\mid \boldsymbol{x}_t] = \boldsymbol{\theta}_t^\top f(\boldsymbol{x}_t)$. 
This corresponds to adjusting the relative importance of each nonlinear effect, while preserving their shapes.
This is achieved by considering the state space model
\begin{align*}
    & \theta_t - \theta_{t-1} \;\sim \mathcal{N}(0,\;Q_t), \\
	& y_t - \theta_t^\top x_t \sim \mathcal{N}(0,\;\sigma_t^2),
\end{align*}
where $\mathcal{N}(\mu, \sigma^2)$ is the multidimensional normal distribution with mean $\mu$ and variance matrix $\sigma^2$, $\theta_t$ is the latent state, $Q_t$ the process noise covariance matrix and $\sigma_t^2$ is the observation variance. Applying the recursive Kalman filter equations as described in section A of \cite{vilmarest2022state} provides us with both $ \theta_t$ and the conditional expectation $\mathbb{E}[y_t\mid \boldsymbol{x}_t]$, which is known to be the best forecast, i.e., minimising the mean square error conditional on past observations and exogenous covariates $x_t$. 
As in \cite{vilmarest2022state}, we run the three variants \textit{Static}, \textit{Dynamic}, and \textit{Viking} of the Kalman filter. The \textit{Static} version is a degenerate case where $Q_t$ is null, which leads to low adaptation. The \textit{Dynamic} variant supposes that $Q_t=Q$ and $\sigma_t=\sigma$ are constants and obtained by grid search optimisation on past observation. Finally, the \textit{Viking} version assumes that $Q_t$ and $\sigma_t$ are updated online (see \cite{vilmarest2022state} for more details).
In Table \ref{table_score_target_agg2}, the GAM model used in the state space models is the one from \cite{obst2021adaptative}, while the
\textit{Static} Kalman filter, \textit{Dynamic} Kalman filter, and \textit{Viking} method are from  \cite{vilmarest2023adaptative}.

\subsubsection{Online aggregation of experts}

Online robust aggregation of experts \citep{Cesa-Bianchi:2006} is a model agnostic technique for time series forecasting. This approach combines various forecasts (called experts) based on their past performance in a streaming manner. 
It allows adaptation to changes in distributions by tracking the best experts.
Sequential expert aggregation assumes that the data are observed sequentially.
The target variable $Y$ (here electricity demand) is supposed to be a bounded sequence, i.e., $Y_1,\dots,Y_T \in [0,B]$, where $B>0$. 
Our goal is to forecast this variable step by step for each given time $t$. At each time $t$, $N$ experts offer forecasts of $Y_t$, denoted by $\left(\hat{Y}_{t}^{1},\dots,\hat{Y}_{t}^N\right) \in [0,B]^N$. 
These experts can be the result of any process, such as a statistical model, a physical model, or human-based expertise.
Then, the aggregation algorithm generates a forecast of $Y_t$ by the weighted average of the $N$ forecasts \[\hat Y_t = \sum_{j=1}^N \hat{p}_{j, t} \;\hat{Y}_{t}^j,\] where the weight $\hat{p}_{j, t} \in \mathbb{R}$ depends on the performance of $\hat Y_t^j$ over the period $\{1, \hdots, t-1\}$. 
Then, $Y_t$ is observed and the next instance starts.

In our study, we run the ML-Poly algorithm, first proposed by \cite{gaillard2014second} and subsequently implemented in \texttt{R} in the \texttt{opera}  package \citep{gaillard2016opera}. This algorithm identifies the best expert aggregation by giving more weight to the experts producing the lowest regret, rendering it noteworthy due to the absence of parameter tuning. 
In Table \ref{table_score_target_agg2}, all the estimators related to data assimilation techniques are combined, i.e., the GAM, the static Kalman filter, the dynamic Kalman filter, and the Viking estimator. 

\subsection{Machine learning}

\subsubsection{Random forests}

Among the most robust machine learning techniques are random forests \cite{breiman2001random}. 
They consist of averaging a given number of decision trees generated by applying classification and regression trees \cite{breiman1984cart} to different subsets of the data obtained by bagging and random sampling of covariates.
Each decision tree estimates the target by a series of logical comparisons on the feature variables. 
An example of decision tree of depth 3 is "if \textit{temperature}  $>$ 30°C, if it is 10 a.m., and if it is a Wednesday, then \textit{electricity demand} = 6 GW". 
Random forest require very little prior knowledge about the problem, which makes them very attractive for benchmarks in applied machine learning problems.
In Table  \ref{table_score_target_agg2}, the random forests all have 1000 trees of depth 6 (the square root of the number of features).
Random forests are usually trained on random subsets of the training sample.
To take advantage of the dependence of samples in time series, the random subsets can be drawn from a given number of consecutive measures. 
This is what is done in the \textit{random forest + bootstrap} architecture \cite{gohery2023random}.

\subsubsection{Gradient boosting}

Gradient boosting \cite{breiman1997arcing,friedman2001greedy} consists in successively fitting the errors of simple models called weak learners, and then aggregating them. 
This is an ensemble technique, like random forests. 
It usually performs better, at the cost of more parameters to calibrate. 
It has demonstrated excellent performance for regression problems \cite{grinsztajn2022tree} and forecasting challenges \cite{makridakis2022m5}.
In tree-based gradient boosting algorithms, weak learners are decision trees, whereas in GAM boosting algorithms \cite{buhlmann2007boosting}, weak learners are spline regression models.

\subsection{Models with mobile-phone data}
As explained in Appendix \ref{app:missing}, the forecasts trained on the dataset "with mobility data" actually consist of two models. 
The first model is trained on the entire dataset "without mobility data". 
The second model estimates the error of the first model on the dataset "with mobility data". 
The GAM "with mobility data" is the sum of the GAM "without mobility data" and of the following GAM
\begin{align*}
    err_{h,t} =\ & \sum_{i=1}^7 \sum_{j=0}^1 \tilde \alpha_{h,i,j}\;\boldsymbol{1}_{\text{DayType}_t=i} \\
    & + f_{h,7}(\text{ToY}_t) +f_{h,8}(\text{Work}_t) +f_{h,9}(\text{Residence}_t)  + \varepsilon_{h,t}. 
\end{align*}
The static Kalman filter, the dynamic Kalman filter, and the Viking estimators "with mobility data" are then computed by summing the effects of the two GAMs.
The GAM boosting "with mobility data" is the sum of the GAM boosting "without mobility data" and of a GAM boosting with all variables (calendar, meteorological, electricity, and mobile phone).
The random forest "with mobility data" is the sum of the \textit{random forest + bootstrap} model "without mobility data" and of a random forest with all variables.
The  \textit{random forest + bootstrap} "with mobility data" is the sum of the \textit{random forest + bootstrap} model "without mobility data" and of a  \textit{random forest + bootstrap} with all variables.

\subsection{Excluding holidays}
As mentioned in Section \ref{sec:benchmark} of the main paper, holidays are known to behave differently from regular days \citep{Krstonijevic2022adaptive}. 
Therefore, we run the same benchmark here, but excluding holidays, as well as the days directly before and after holidays, from both training and testing.
Table \ref{table_score_target_agg} shows that, when excluding holidays, incorporating mobility data improves the best performance (aggregation of experts) by 8\% in RMSE and 6\% in MAPE.
Once again, the global order of magnitude of the performance gains across all models is 10 $\%$.
Note that these gains are significant, because they leave the confidence interval obtained by bootstrapping.
\begin{table}[ht]
\centering
\caption{\textbf{Benchmark excluding holidays.}} 
\begin{tabular*}{\textwidth}{@{\extracolsep\fill}lcccc}
  \toprule
 & \multicolumn{2}{@{}c@{}}{ Without mobility data  } & \multicolumn{2}{@{}c@{}}{With mobility data } \\\cmidrule{2-3}\cmidrule{4-5}%
 & RMSE (GW) & MAPE (\%)  &  RMSE (GW) & MAPE (\%) \\
  \midrule
  \textit{Model} &&&&\\
  Persistence (1 day) & 4.0 \; $\pm$  0.2 \; & 5.0 \; $\pm$ 0.3 \;  & N.A. & N.A.\\
    SARIMA  & 2.0 \; $\pm$  0.2  \; & 2.6 \; $\pm$ 0.2 \;  & N.A. & N.A. \\
  GAM & 1.70 $\pm$ 0.06 & 2.6\;\; $\pm$ 0.1\;\; & \textbf{1.55} $\pm$ 0.05 & \textbf{2.43} $\pm$ 0.08 \\
  \midrule
  \textit{Data assimilation technique} &&&&\\
  Static Kalman filter & 1.43 $\pm$ 0.05 & 2.20 $\pm$ 0.08 & 1.07 $\pm$ 0.04 & 1.63 $\pm$ 0.06 \\
  Dynamic Kalman filter & 1.10 $\pm$ 0.04 & 1.58 $\pm$ 0.05  & 0.96 $\pm$ 0.03 & 1.39 $\pm$ 0.04\\
    Viking & 0.98 $\pm$ 0.04 & 1.33 $\pm$ 0.04 & 0.98 $\pm$ 0.03 & 1.41 $\pm$ 0.05\\
    Aggregation of experts & 0.96 $\pm$ 0.04 & 1.36 $\pm$ 0.04  & \textbf{0.88} $\pm$ 0.03  & \textbf{1.28} $\pm$ 0.04\\
    \midrule
    \textit{Machine learning}\\
    GAM boosting & 2.3  $\pm$ 0.1 & 3.3  $\pm$ 0.2 & 2.2 $\pm$ 0.1 & 3.1 $\pm$ 0.2 \\
    Random forests & 2.1 $\pm$ 0.1 & 3.0 $\pm$ 0.1 & \textbf{1.8} $\pm$ 0.1 & \textbf{2.4} $\pm$ 0.1\\
    Random forests + bootstrap & 1.9 $\pm$ 0.1 & 2.6 $\pm$ 0.1 &\textbf{1.8} $\pm$ 0.1 & \textbf{2.4} $\pm$ 0.1\\
   \bottomrule
\end{tabular*}
\label{table_score_target_agg}
\end{table}

\section{Change point detection}
In this appendix, we detail and justify the use of the model in Section \ref{sec:energy_savings}, as well as the change point detection algorithm applied on top of it.
\subsection{Model for seasonality}
The model used in Section \ref{sec:energy_savings} to capture the dependence of calendar and meteorological data on electricity demand is the following direct adaptation of the GAM of \cite{obst2021adaptative}
\begin{align*}
    \mathrm{Load}_{h,t} =\ & \sum_{i=1}^7 \sum_{j=0}^1 \alpha_{h,i,j}\;\boldsymbol{1}_{\text{DayType}_t=i}\;\boldsymbol{1}_{\text{DLS}_t=j} \\
    & + f_{h,1}(\text{ToY}_t) + f_{h,2}(\text{Temp95}_{h,t}) + f_{h,3}(\text{Temp99}_{h,t}) \\
    & + f_{h,4}(\text{TempMin99}_{h,t},\; \text{TempMax99}_{h,t})  + \varepsilon_{h,t}. 
\end{align*}
Notice that this corresponds to removing the dependence in the timestamp $t$ and in the lags Load1D and Load1W from equation \eqref{eq:modele_GAM_FR}.
On the one hand, these features were removed because they only capture the trend of the signal without explaining the phenomena at stake, which interferes with the interpretability of the model.
On the other hand, the remaining features account for well-known repeated phenomena, knowingly the effects of weekends  (in DayType), of holidays (in ToY), and of heating and cooling (in the temperature smoothings).
They explain the seasonality of the signal.
This GAM is trained from 2014-01-01 to 2018-01-01. 
The residuals $\text{res} = \text{Load} - \hat{\text{Load}}$ are then evaluated from 2018-01-01 to 2023-03-01.
Between 2018-01-01 and 2020-01-01, this GAM has an average MAPE of $2.1 \%$ and an average RMSE of $1.6$ GW.
This is comparable to the performance of the GAM of \cite{obst2021adaptative}, which has an average MAPE of $1.6 \%$ and an average RMSE of $1.2$ GW.
At the cost of a slightly lower performance, this GAM is more explainable because it only takes seasonal phenomena into account.
It is therefore a good model to forecast what electricity demand should be over a multi-year time horizon, assuming  that the electricity consumption behaviour will stay  unchanged. 

\subsection{Descriptive analysis of residuals}
\begin{figure}
    \centering
    \caption{\textbf{Descriptive statistics of the residuals.}}
    \includegraphics[width = 0.49\textwidth]{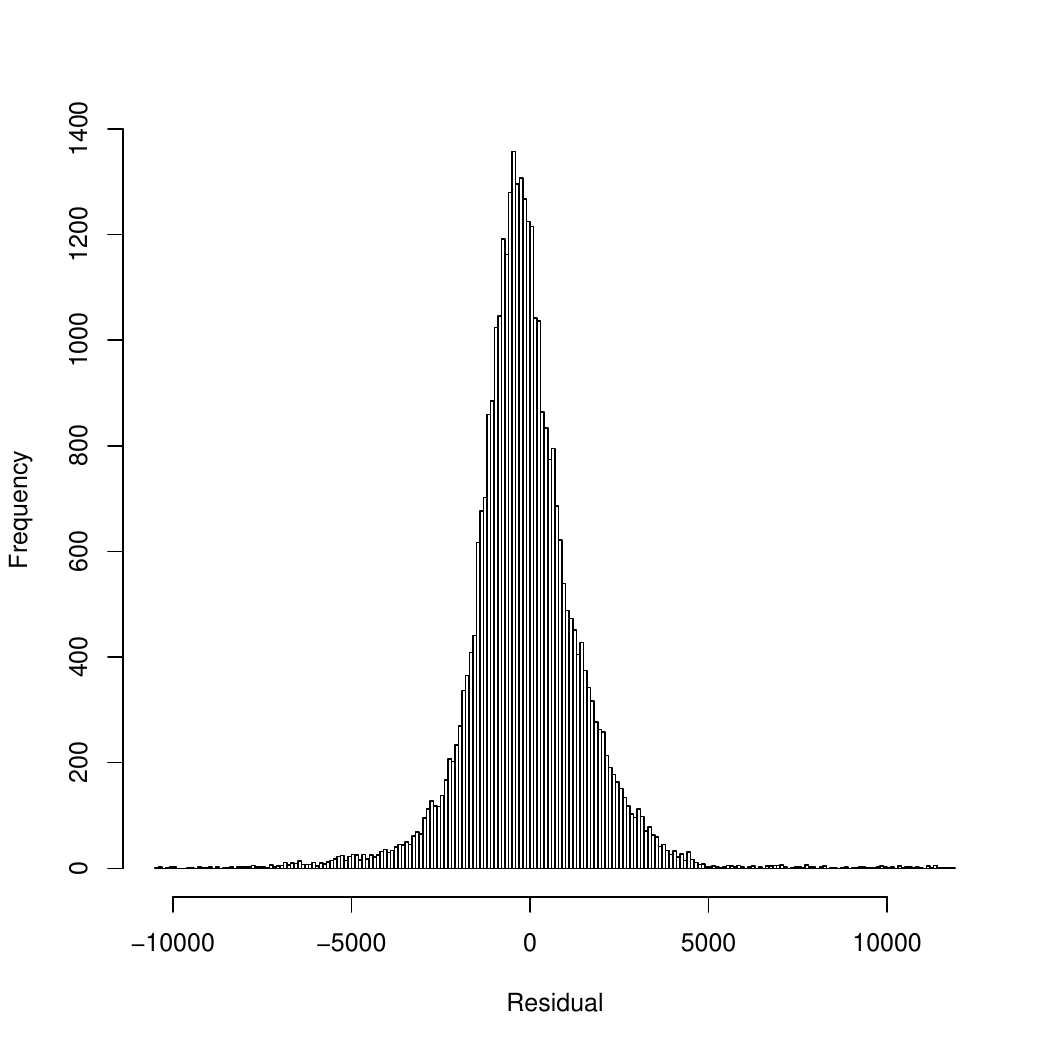}
    \includegraphics[width = 0.49\textwidth]{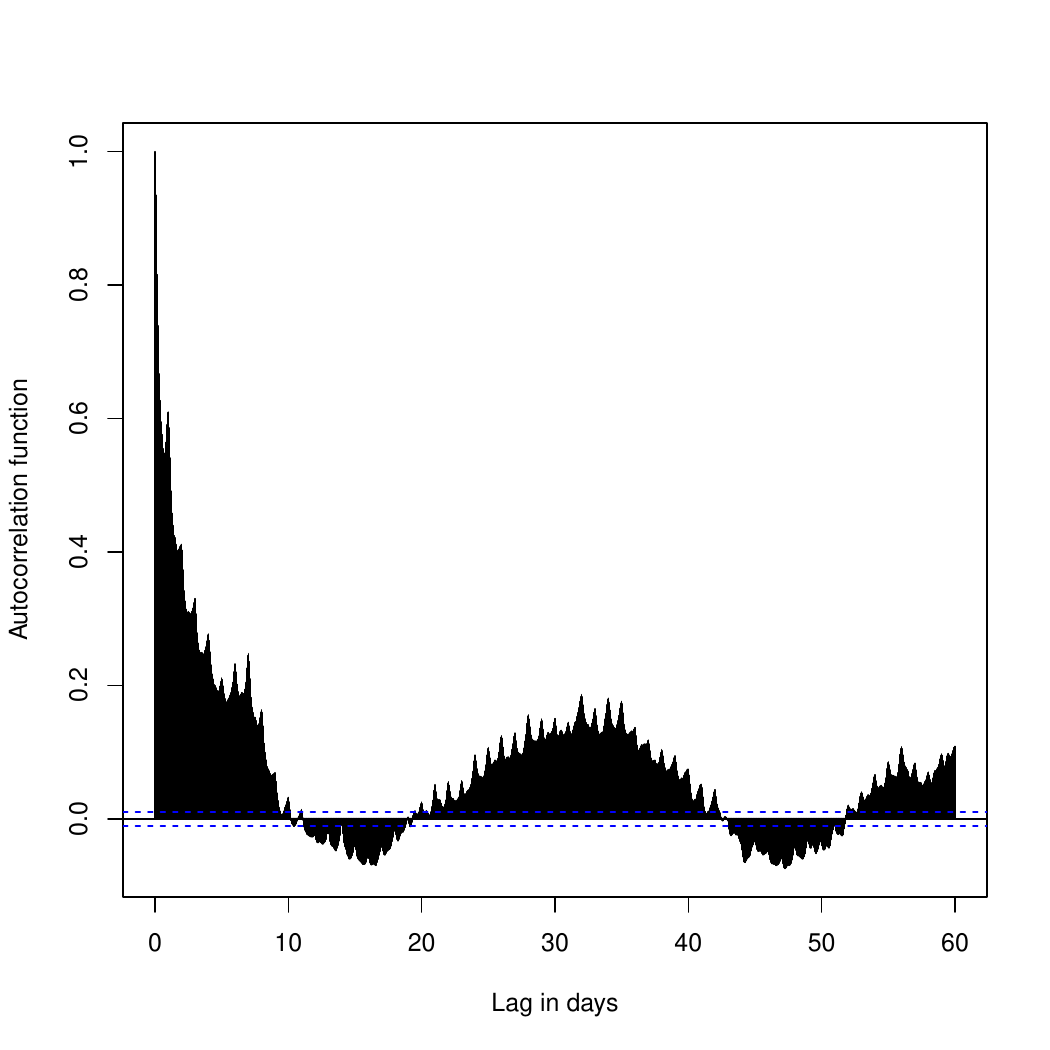}
    \flushleft
    {\small \textbf{Left:} Histogram of the residuals between 2018-01-01 and 2020-01-01. \textbf{Right:} Autocorrelation function of the residuals between 2018-01-01 and 2020-01-01. Dotted blue lines draw a confidence interval for the precision of the auto-correlation estimators.}
    \label{fig:description_residuals}
\end{figure}

In this paragraph, we focus on the period spanning from 2018-01-01 to 2020-01-01.
As shown in Figure \ref{fig:description_residuals} (left), the residuals histogram presents a bell shape.
Since we have $2 \times 365 \times 48 = 35040$ observations, we chose the number of breaks in the histogram to be $\lfloor \sqrt{35040} \rfloor = 187$. 
The T-test reveals that the expectation of the residuals is significantly lower than zero (the p-value is lower than $2.2 \times 10^{-16}$) and that it lies in the interval of $[-0.16\;  \text{GW},\; -0.12\; \text{GW}]$ with 95 \% confidence.
The empirical mean is $-0.14$ GW, while the empirical standard deviation is $1.6$ GW.
The Anderson-Darling normality test shows that the residuals do not follow a normal law (the p-value is less than $2.2 \times 10^{-16}$).
Moreover, as shown in Figure \ref{fig:description_residuals} (right), the autocorrelations of the residuals decrease slowly  and are significantly higher than zero, suggesting that the residuals are not stationary.
Indeed, this is confirmed by running the Box-Ljung test with a 1-day window (the p-value is less than  $2.2 \times 10^{-16}$).
Both the facts that the expectation of the residuals is significantly lower than zero and that the residuals are not stationary indicate that other phenomena than calendar seasonality and temperature are involved, although their impact is moderate as the estimator performs well.
This suggests that, even in this period without major breaks such as COVID-19 or sobriety, other features are useful to understand the electricity demand.

\subsection{Ranking changes in the data distribution }
The descriptive analysis shows that the residuals are not stationary. 
Therefore, from a statistical point of view, it is pointless to look for the ruptures observed in Figure \ref{fig:obst} in absolute terms.
In fact, the more precise the technique for detecting change points becomes, the more change points will be detected everywhere.
This is why we need quantitative information about the importance of the change points in order to rank them and determine which are the most significant change points.
Indeed, several metrics have been developed to measure  the importance of change points \cite{Aminikhanghahi2017a}.
To assess the significance of the number of change points, we sequentially compare the standard deviation of the residuals with the amplitude of the change points.
This results in 10 change points being considered in the following analysis.
The principle behind offline changing in mean techniques is to segment the signal in such a way that approximating the signal to its mean at each segment results in the lowest possible variance.
However, finding such an optimum is computationally expensive.
Therefore, faster algorithms have been developed to find approximations to the optimal change points, such as the binary segmentation algorithm used in Figure \ref{fig:obst}.

\section{Statistical analysis}
In this appendix, we complement the variable selection detailed in the \textcolor{blue}{Results section} 
to further justify the study of the \textit{work} index in the statistical analysis of Section \ref{sec:stat_ana}.

\subsection{Variable selection: Hoeffding D-stastics and Shapley values}
\label{sec:variable_selection}
This paragraph complements the mRMR variable ranking performed in the \textcolor{blue}{Results section}. 
To examine the variable selection process more closely, we compute the Hoeffding D-statistic, as shown in Table \ref{table:hoeff}.
It is a distribution-free measure of the dependence between variables \cite{hoeffding1948a}. 
The closer it gets to 1, the greater the dependence.  
We then compute the Shapley values of the same variables using the \texttt{SHAFF} algorithm \cite{pmlr-v151-benard22a}, as shown in Table \ref{table:shapley}.
Notice that, with the three ranking methods, the 3 most important variables, in order of importance, are the \textit{temperature}, the \textit{work} index, and then the \textit{time of year}.
Interestingly, the effect of the work index only becomes clear after correcting the electricity demand from the temperature dependence.
Notice how the importance of \textit{tourism} and \textit{time of year} decreases when correcting electricity demand for temperature, due to their high correlation with temperature. 
As a result of this analysis, the \textit{tourism}  and \textit{residents} indices do not seem to have a significant impact on the French electricity demand.
\begin{table}[ht]
\centering
\caption{\textbf{Hoeffding D-statistic.} } 
\begin{tabular*}{\textwidth}{llllllll}
  \toprule
 &    Temp95 &  Work & Residence & Tourism & Toy & Dow  & Holidays \\
  \midrule
  Load &   \textbf{0.29}& 0.035 & 0.093& 0.19 & 0.069 & 0.010& 0.000071\\
  Load $\backslash$ Temp & 0.018& \textbf{0.18}& 0.010 & 0.025 & 0.017& 0.091&0.00066 \\
  Load  $\backslash$(Temp,Work) & 0.0057 & 0.0067 & 0.014 & 0.011 & \textbf{0.038} &  0.0057 & -0.00001\\
   \bottomrule 
\end{tabular*}
\vspace{1em}
\flushleft
{\small The statistic is computed using all available days from 2019-07 to 2022-03. \textit{Load} $\backslash$ (\textit{features}) stands for the \textit{Load} corrected for the effects of the \textit{features}.}
\label{table:hoeff}
\end{table}

\begin{table}[ht]
\centering
\caption{\textbf{Shapley values.}} 
\begin{tabular*}{\textwidth}{llllllll}
  \toprule
 &    Temp95 &  Work & Residence & Tourism & Toy & Dow  & Holidays \\
  \midrule
  Load & \textbf{0.31}  & 0.05 & 0.06  & 0.14 & 0.21& 0.03 & 0.002 \\
   & $\pm$ 0.05  & $\pm$ 0.02 & $\pm$ 0.03 & $\pm$ 0.01 & $\pm$ 0.01 & $\pm$ 0.02 & $\pm$ 0.003\\
   &&&&&&\\
  Load $\backslash$ Temp & 0.041 & \textbf{0.26}& 0.035& 0.072& 0.11& 0.19& 0.04\\
   & $\pm$ 0.007& $\pm$ 0.01& $\pm$ 0.005& $\pm$0.005& $\pm$ 0.02& $\pm$ 0.02& $\pm$ 0.01\\
   &&&&&&\\
  Load  $\backslash$ (Temp, Work) & 0.11& 0.10& 0.06& 0.07& \textbf{0.28}& 0.04& 0.003\\
  & $\pm$ 0.01 & $\pm$ 0.01 & $\pm$ 0.01 & $\pm$ 0.01 & $\pm$ 0.01 & $\pm$ 0.01& $\pm$ 0.008\\
   \bottomrule 
\end{tabular*}
\vspace{1em}
\flushleft
{\small Shapley values are computed on all available days from 2019-07 to 2022-03. \textit{Load} $\backslash$ (\textit{features}) stands for the \textit{Load} corrected for the effects of the \textit{features}.}
\label{table:shapley}
\end{table}

\subsection{Work index and calendar features}
\label{sec:remote_wkg}
 
The variable selection analysis in Appendix \ref{sec:variable_selection} shows that the \textit{work} indicator has a very strong effect on the electricity demand, being the second most explanatory variable.
To better understand this effect, we compare in Table \ref{table_GAM} the performance of GAMs where we progressively add the features in the order of importance suggested by the variable selection analysis.
The \textit{Temp} GAM corresponds to the model
\[\text{Load}_{h,t} = f_{h}(\text{Temp95}_{h,t}) + \varepsilon_{h,t}.\]
The \textit{Temp + Work} GAM corresponds to the model
\[\text{Load}_{h,t} = f_{h, 1}(\text{Temp95}_{h,t}) + f_{h, 2}(\text{Work}_{h,t})  + \varepsilon_{h,t}.\]
The \textit{Temp + Time} GAM corresponds to the model
\begin{align*}
    \text{Load}_{h,t} =\ & \sum_{i=1}^7 \sum_{j=0}^1 \alpha_{h,i,j}\;\boldsymbol{1}_{\text{DayType}_t=i}\;\boldsymbol{1}_{\text{DLS}_t=j} \nonumber\\
    & + \beta \boldsymbol{1}_{\text{Holidays}_t} + f_{h,1}(\text{ToY}_t) + f_{h,2}(\text{Temp95}_{h,t})  + \varepsilon_{h,t}.
\end{align*}
The \textit{Temp + Time + Work} GAM corresponds to the model
\begin{align*}
    \text{Load}_{h,t} =\ & \sum_{i=1}^7 \sum_{j=0}^1 \alpha_{h,i,j}\;\boldsymbol{1}_{\text{DayType}_t=i}\;\boldsymbol{1}_{\text{DLS}_t=j} \nonumber\\
    & + \beta \boldsymbol{1}_{\text{Holidays}_t} + f_{h,1}(\text{ToY}_t) + f_{h,2}(\text{Temp95}_{h,t}) +  f_{h, 3}(\text{Work}_{h,t})  + \varepsilon_{h,t}.
\end{align*}
The \textit{Temp +  Work + Lags} GAM corresponds to the model
\begin{align*}
    \text{Load}_{h,t} =\ & f_{h, 1}(\text{Temp95}_{h,t}) + f_{h, 2}(\text{Work}_{h,t}) + \sum_{i=1}^7  \alpha_{h,i,j}\;\boldsymbol{1}_{\text{DayType}_t=i}\;\text{Load1D}_{h,t} \nonumber\\
    & + \beta \text{Load1W}_{h,t}  + \varepsilon_{h,t}.
\end{align*}
The \textit{Temp +  Time + Lags} GAM corresponds to the model
\begin{align*}
    \text{Load}_{h,t} =\ & \sum_{i=1}^7 \sum_{j=0}^1 \alpha_{h,i,j}\;\boldsymbol{1}_{\text{DayType}_t=i}\;\boldsymbol{1}_{\text{DLS}_t=j} + \beta \boldsymbol{1}_{\text{Holidays}_t} + f_{h,1}(\text{ToY}_t)  \\
    &+ f_{h,2}(\text{Temp95}_{h,t})+ \sum_{i=1}^7 \gamma_{h,i,j}\;\boldsymbol{1}_{\text{DayType}_t=i}\;\text{Load1D}_{h,t}\\
    &+ \lambda \text{Load1W}_{h,t}  + \varepsilon_{h,t}.
\end{align*}
The \textit{All variables} GAM corresponds to the model
\begin{align*}
    \text{Load}_{h,t} =\ &\sum_{i=1}^7 \sum_{j=0}^1 \alpha_{h,i,j}\;\boldsymbol{1}_{\text{DayType}_t=i}\;\boldsymbol{1}_{\text{DLS}_t=j} + \beta \boldsymbol{1}_{\text{Holidays}_t} + f_{h,1}(\text{ToY}_t)  \\
    &+ f_{h,2}(\text{Temp95}_{h,t})+ \sum_{i=1}^7 \gamma_{h,i,j}\;\boldsymbol{1}_{\text{DayType}_t=i}\;\text{Load1D}_{h,t} + \lambda \text{Load1W}_{h,t}\\
    &+ f_{h,3}(\text{Work}_{h,t}) + \varepsilon_{h,t}.
\end{align*}
The p-values of the Fisher tests assessing the significance of the GAM effects are less than 5 $\%$ for all the GAMs.
Notice on Table \ref{table_GAM} how replacing calendar data by the work index is beneficial during atypical events which behaviour differs from the past, i.e., the \textit{sobriety} period.
Indeed, the time variables are only relevant during the \textit{normal period} spanning from July 2023 to September 2023, during which they still benefit from the work index. 
During the \textit{sobriety} period, the time variables ---which only reconstruct past behaviour--- are less powerful than the work index, which does not benefit from being coupled with them.

\begin{table}[ht]
\centering
\caption{\textbf{Integration of mobility data in GAMs.} } 
\begin{tabular*}{\textwidth}{@{\extracolsep\fill}lccccc}
  \toprule
 &  \multicolumn{2}{@{}c@{}}{ Normal period  } & \multicolumn{2}{@{}c@{}}{Sobriety} \\\cmidrule{2-3}\cmidrule{4-5}%
 & RMSE (GW) & MAPE (\%)  &  RMSE (GW) & MAPE (\%) \\
  \midrule
  \textit{Baseline} &&&&&\\
  Persistence (1 day) & 3.49   & 5.36  & 4.03 & 5.26\\
  \midrule
    \textit{GAM} \\
  Temp & 3.53& 6.78& 6.13& 9.60\\
  Temp + Work & 1.62& 3.10& 4.98& 8.11 \\
  Temp +  Time & 1.33& 2.46& 5.60& 9.62\\
  Temp + Time + Work & 1.09&2.02& 5.24& 9.00\\
  Temp  + Work + Lags& 1.11 & 1.92& \textbf{2.11}& \textbf{3.13} \\
  Temp  +  Time + Lags& 0.89&1.52& 2.61& 4.29\\
  All variables & \textbf{0.80} & \textbf{1.38}& 2.60& 4.32\\
   \bottomrule
\end{tabular*}
\vspace{1 em}
\flushleft
This benchmark covers all days, including holidays.
\label{table_GAM}
\end{table}

\subsection{Work dynamics}
\begin{figure}
    \centering
    \caption{\textbf{Effect of the work index on a given day at a given hour.}}
    \includegraphics[width = 0.49\textwidth]{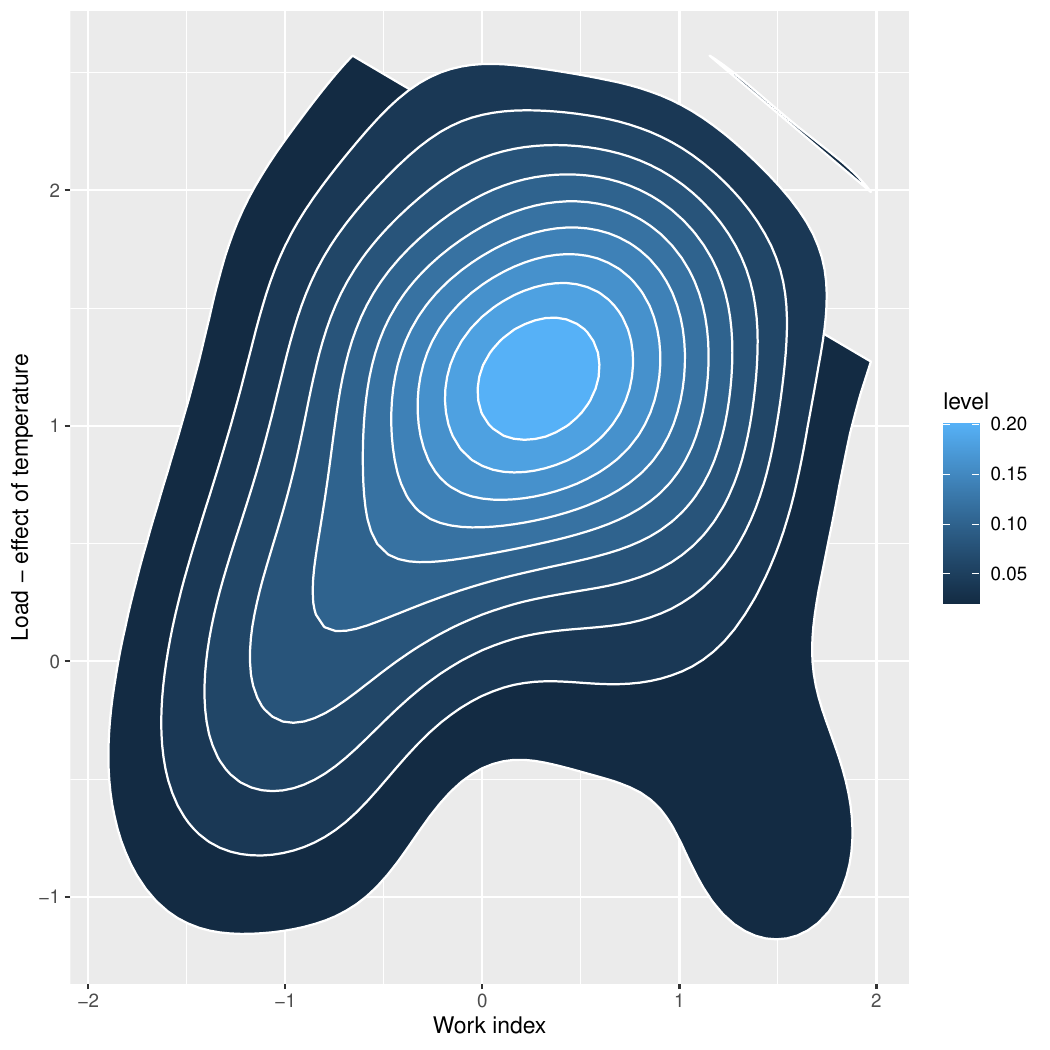}
    \includegraphics[width = 0.49\textwidth]{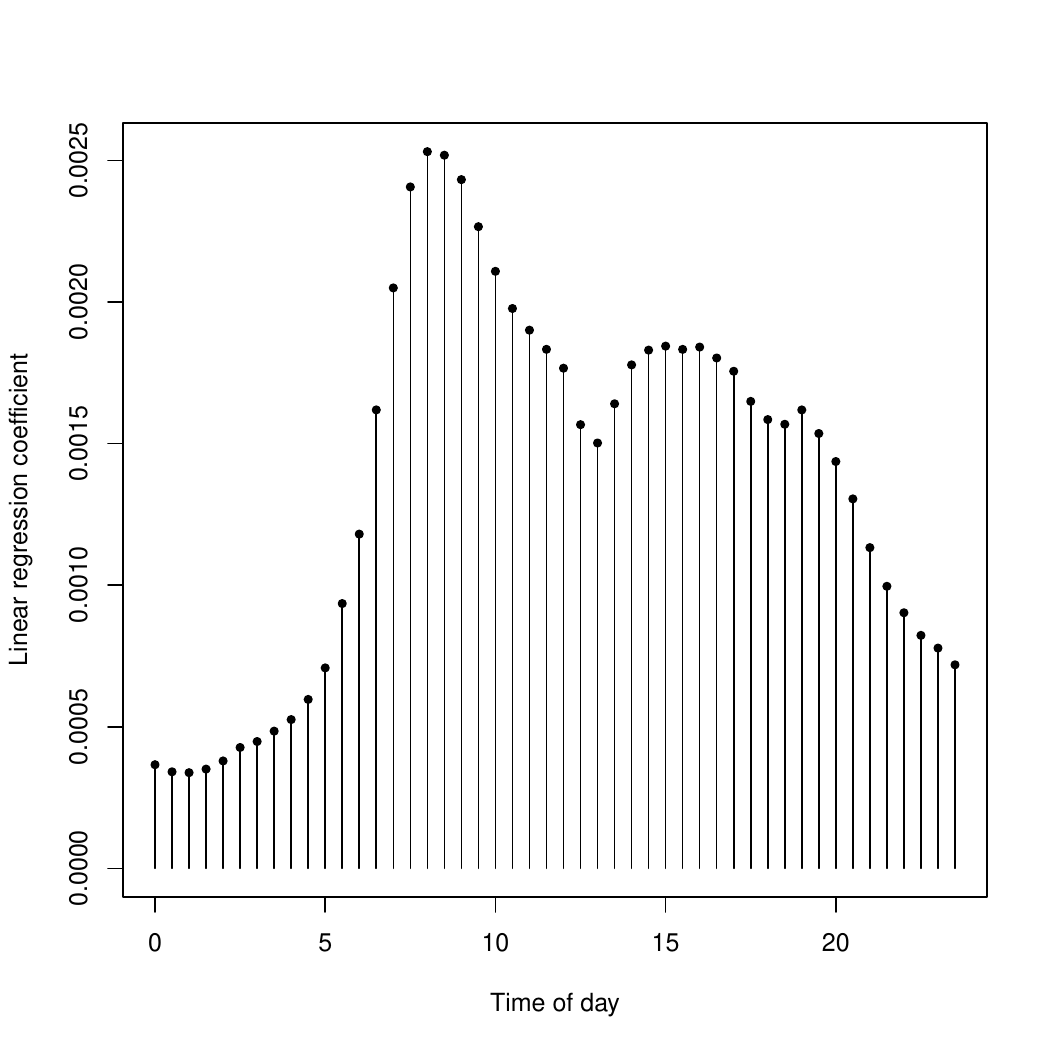}
    \flushleft
    {\small \textbf{Left:} 2d density plots of residuals as function of the work index at 10 am on Wednesdays between 07-2019 and 03-2022. \textbf{Right:} Regression coefficient of the work index on electricity demand corrected for the effect of temperature on the training set spanning from 07-2019 to 03-2022.}
    \label{fig:wednesday}
\end{figure}

In Section \ref{sec:stat_ana}, we explained how the \textit{work} index captures the effects of both the \textit{day of week} and the \textit{holidays} features. 
However, in both Section \ref{sec:benchmark} and Appendix \ref{sec:remote_wkg}, we showed that the \textit{work} index improves the performance of the forecast, beyond the effect of the calendar features.
The aim of this paragraph is to study this effect. 
Therefore, to remove the effects of the \textit{time of day} and of \textit{holidays}, we work on a specific day (here Wednesday) and by removing the holidays.
Figure \ref{fig:wednesday} (left) shows how the electricity demand on Wednesdays is still positively influenced by the work index.
Furthermore, as expected, Figure \ref{fig:wednesday} (right) shows that the effect of the \textit{work} index is more important during working hours (from 6 a.m. to 8 p.m.).
These results confirm that on Wednesdays a high \textit{work}  index corresponds to a high electricity demand.
This effect could be due to economic growth (a higher economic activity corresponding to both more people working which raises the \textit{work} index, and to a higher electricity demand) and to the energy saving due to remote working (a lower office occupancy corresponding both to a lower \textit{work} index and to a lower electricity demand).
\end{appendices}

\end{document}